\documentclass[prd,amsmath,preprintnumbers,superscriptaddress,floatfix,letterpaper,showpacs,english]{revtex4}
\usepackage{amssymb}
\usepackage{amsmath}
\usepackage{amsfonts}
\usepackage{epsfig}
\usepackage{graphicx}
\usepackage{tabularx}
\usepackage{latexsym}
\usepackage{subfigure}
\usepackage{verbatim}
\usepackage{color}
\usepackage{braket}
\usepackage{multirow}
\usepackage{dcolumn}
\usepackage{mathtools}
\usepackage{placeins}
\usepackage{epstopdf}
\usepackage{wrapfig}
\usepackage{hyperref}
\usepackage{diagbox}
\usepackage{booktabs}
\usepackage{simplewick}
\usepackage[normalem]{ulem}
\usepackage[latin1]{inputenc}
\usepackage{extarrows}
\usepackage{lipsum,babel}
\usepackage[toc,page]{appendix}

\def\gev{\mathrm{GeV}}

\def\lsim{\raise0.3ex\hbox{$<$\kern-0.75em\raise-1.1ex\hbox{$\sim$}}}
\def\gsim{\raise0.3ex\hbox{$>$\kern-0.75em\raise-1.1ex\hbox{$\sim$}}}

\newcommand{\bee}{\begin{equation}}
\newcommand{\ee}{\end{equation}}

\renewcommand\sout{\bgroup \color{red} \ULdepth=-.5ex \ULset}

\newcommand{\half}{ {\textstyle\frac{1}{2}} }
\renewcommand{\slash}[1]{#1 \hspace{-0.45em} / }
\usepackage{mathtools}
\DeclarePairedDelimiter\abs{\lvert}{\rvert}%
\allowdisplaybreaks[1] 
\newcommand{\bm}[1] {\mbox{\boldmath{$#1$}}}

\def\eq{\begin{eqnarray}}
\def\en{\end{eqnarray}}
\def\ep{\epsilon}

\def\la{\lambda}

\begin{document}
\preprint{}


\title{$\rho$ meson unpolarized generalized parton distributions \\ with a light-front constituent quark model }

\author{Bao-Dong Sun}
\email{sunbd@ihep.ac.cn}
\affiliation{Institute of High Energy Physics, Chinese Academy of Sciences, Beijing 100049, People's Republic of China}
\affiliation{School of Physics, University of Chinese Academy of Sciences, Beijing 100049, People's Republic of China}
\author{Yu-Bing Dong}
\affiliation{Institute of High Energy Physics, Chinese Academy of Sciences, Beijing 100049, People's Republic of China}
\affiliation{School of Physics,  University of Chinese Academy of Sciences, Beijing 100049, People's Republic of China}
\affiliation{Theoretical Physics Center for Science Facilities (TPCSF), CAS, Beijing 100049, People's Republic of China}
\noaffiliation

\begin{abstract}
	
We study $\rho$ meson unpolarized generalized parton distributions based on a light-front constituent quark model
where the quark-antiquark-meson vertex is constructed under the symmetric loop momentum convention. The form factors
and some other low-energy observables of the $\rho$ meson are calculated.  Moreover, the contributions to the form
factors and generalized parton distributions from the valence and nonvalence regimes are discussed and analyzed in detail.
In the forward limit, the usual structure functions are estimated as well.
In addition, by evolving the moments of the obtained structure functions to the scale of the lattice calculation, we give the factorization scale of our quark model. It is found that the present phenomenological model is reasonable to
describe the general properties of $\rho$ meson.
\end{abstract}
\date{\today}
\maketitle

\section{Introduction}

It is known that the usual parton distribution functions (PDFs) and electromagnetic form
factors (FFs) shed light on the ``one-dimensional" structure of hadrons~\cite{Marukyan2015}.
Moreover, generalized parton distributions (GPDs) naturally embody the information
of both PDFs and FFs, and therefore they display the unique properties to present a ``(3D)" 
description for the transverse and longitudinal partonic degrees of freedom inside the
system, and they contain promising potential which gives arise to ideals of ``quark/gluon imaging" of
hadrons~\cite{Marukyan2015}. Many theoretical investigations have been carried out on the general
properties of GPDs for a hadronic system~\cite{Diehl2003}. It is believed that the studies of
GPDs are closely related to the processes of deeply virtual Compton scattering and  the deeply
virtual meson electroproduction~\cite{Ji2006,Kumericki2016,Goeke2001}.
By comparing with experimental measurements, one can obtain  possible constrains on the GPDs of a hadron~\cite{Aidala2013,Airapetian2017}. With the help of  sum rules, the unpolarized GPDs are directly connected
to the electromagnetic FFs of the system. There are some empirical parametrizations for GPDs. For the
nucleon case, those parameterizations can be obtained by fitting the experimental data to the Dirac, Pauli, and axial
FFs ~\cite{Kirchner2003,Guidal2004nd,Diehl2013Feb19,Selyugin2015Oct12,Sharma2016}. In the forward limit,
GPDs reproduce the usual PDFs, and thus a description of GPDs can also be built with the help of the experimental data
of PDFs~\cite{Selyugin2015Oct12}. In addition, the moments of GPDs can provide other new information as well, such as
the neutron asymmetry ~\cite{Zhang2015} and the quark orbital angular momentum~\cite{Hoodbhoy1998,Ji2016Dec7}.\\

Many endeavors have been made to study the GPDs of simple hadrons in the literature, like the studies of a pion~\cite{Broniowski2011,Choi2001,Fanelli:2016aqc,Mezrag2016,Kumericki2016};
of a proton and neutron~\cite{Kroll2017,Pire2009,Diehl2013Feb19,Selyugin2015Oct12,Sharma2016,Rinaldi2017}, and of the light nuclei, ${}^3$He~\cite{Rinaldi2013,Zhang2015} or deuteron~\cite{Berger2001,Kirchner2003,Cano2004,Dong2013,Mondal2017}. In those works, different approaches have been employed. They include the chiral quark models employing the Nambu-Jona-Lasinio model, the spectral quark model~\cite{Broniowski2003,Broniowski200878,Broniowski2011}, the covariant constituent quark models (CCQMs)~\cite{Choi2001,Frederico2009,Fanelli:2016aqc},
the Dyson-Schwinger equation approach~\cite{Mezrag2016}, the AdS/QCD inspired light-front wave functions~\cite{Rinaldi2017}, and some empirical parametrizations as already
mentioned above. Among those phenomenological approaches, the light-cone constituent quark model (LCCQM), one
of the CCQMs, is a quite suitable and successful approach to be applied for the studies of the quark-hadron vertex
and of the hadron properties, as has been pointed out by Refs.~\cite{Brodsky1998,Fanelli:2016aqc}. Besides the various model-dependent studies, some lattice QCD calculations have been also performed~\cite{Dalley2003Jul21,Hagler2010}. It is believed that those lattice simulations, together with the experimental data,
can be employed to check and make a judgement for the different phenomenological models. \\

Apart from the pion (spin-0) and nucleon (spin-1/2) targets, the deuteron (spin-1) target is also common experimentally. The GPDs of a deuteron have been already defined through the matrix operators on the light front~\cite{Berger2001}, and the partonic structures and FFs of the deuteron have been formally explored
through different approaches as well~\cite{Cano2004,Dong2013,Mondal2017,Cosyn:2017ekf,Dong:2008mt,Liang:2015zba}. We know that the deuteron is a weakly bound
system of a proton and a neutron and approximately satisfies the isospin symmetry. Therefore, by considering
the GPDs of the proton and neutron, one may obtain the information of the deuteron GPDs~\cite{Kirchner2003,Cano2004, Airapetian2010}.\\

The $\rho$ meson, which is a spin-1 particle as well, is usually regarded as a $q\bar{q}$ bound state in CCQMs. Some lattice results~\cite{Glozman2011} have already shown that the $\rho$ meson is approximately a pure
${}^3S_1$ state with only $\sim~1\%$ admixture of the ${}^3D_1$ wave, and, consequently, in the rest frame, the valence quarks carry out almost completely the spin of the $\rho$ meson. This conclusion provides a solid support to employ the constituent quark model to explore the $\rho$ meson structure as a pure $q\bar{q}$ system. It should be stressed that the most previous studies of the $\rho$ meson focus on its FFs~\cite{Melo1997,Cardarelli1995,Aliev2004,Choi2004,Biernat2014,Melo2016,Krutov:2016uhy,He:2004ba}. The only one lattice QCD calculation for the moments of the
unpolarized $\rho$ meson PDFs appeared two decades ago~\cite{Best1997}, which was performed at the scale $Q= 2.4~\gev$. With a quenched approximation, Ref.~\cite{Best1997} obtained the $n$th moments of its structure functions which, is meaningful only when comparing with the nonsinglet valence quark distributions. Later on, the result of QCD sum rules for the $\rho$ meson structure functions in Ref.~\cite{Oganesian2001} matches the lattice calculation well.
As for the $\rho$ GPDs, there are some possible indirect approaches to access them, such as through the connection with generalized distribution amplitudes, via the double distributions~\cite{Diehl:1998dk,Diehl2003,Anikin:2005ur} or the Radon transformation~\cite{Teryaev:2001qm}. Thus, it is of a great interest to see what the GPDs of the $\rho$ meson look like with the help of the LCCQM model. This study may
be even useful to understand the processes involving the $\rho$ meson lepton production such as $e+N\rightarrow e+\rho^0+N $~\cite{Mankiewicz1998,Cano2004,Morrow2009,Airapetian2017} or the process of $\gamma\gamma*\rightarrow\rho\rho$~\cite{Anikin:2003fr} and the future Electron-Ion Collider(EIC)  experiments~\cite{Boer:2011fh,Accardi:2011mz}.\\

In analogy to the deuteron case, we introduce the GPDs of the $\rho$ meson and apply the LCCQM for the study of its
unpolarized GPDs. Particularly, the GPDs with different skewness $\xi$ will be discussed in detail. It should be mentioned that in the LCCQM, the separation of the valence (Dokshitzer- Gribov-Lipatov-Altarelli-Parisi, DGLAP) and nonvalence (Efremov-Radyushkin-Brodsky-Lepage, ERBL) regimes is transparent after the integration over the poles in the Dirac propagators of $k^-=k^0-k^3$, i.e. the minus component of the loop momentum. Consequently, we can further study the contributions to the properties of the $\rho$ meson, like its FFs and GPDs, from the valence and nonvalence regimes at different values of $\xi$. \\

This paper is organized as follows. Section~\ref{sec:GPDs_for_hadrons_with_Spin_1} gives a brief introduction to the
general decomposition of GPDs for the spin-1 $\rho$ meson. Section~\ref{sec:The_model} shows a description of the LCCQM. Moreover, in Sec.~\ref{sec:Results}, we display the main numerical results for the $\rho$ meson FFs and its unpolarized GPDs with the LCCQM. In addition, in Sec.~\ref{sec:evolution}, we discuss the QCD evolution of the moments of the $\rho$ meson PDFs and make a comparison to the lattice calculation. Finally, Sec.~\ref{sec:summary} is devoted to a short summary.

\section{GPDs for hadrons with Spin 1}
\label{sec:GPDs_for_hadrons_with_Spin_1}

The notations in this work are~\cite{Frederico2009}
\eq
t &=&\Delta^2=(p'-p)^2 \ , \nonumber \\
\xi &=&-\frac{\Delta\cdot n}{2P\cdot n}= -\frac{\Delta^+}{2P^+} \ , \ \ \abs{\xi}=\frac{\Delta^+}{2P^+} \ , \ \ \ \ \ (\,|\xi\,|\le1) \nonumber \\
x &=&\frac{k\cdot n}{P\cdot n}=\frac{k^+}{P^+} \ , \ \ \ \ \ \ (-1\le x\le1) \ ,
\en
where $p$ and $p'$ are the 4-momenta of the incoming and outgoing $\rho$ mesons, $P=({p'+p})/{2}$, $\Delta=p'-p $,
$n$ is a lightlike 4-vector with $n^2=0$, and $k$ is the 4-momentum in the loop which will be specified in next
section.
The skewness variable $\xi$ plays a similar role as the Bjorken variable~\cite{Ji1997,Airapetian2010}. \\

The helicity counting rules restrict that there are totally nine helicity conserving GPDs of the spin-1 particle
for each quark flavor and the gluons. Five of them are unpolarized (averaged over helicities), and the other four are
polarized (sensitive to helicities). The helicity-averaged GPDs are defined through the two-parton correlation function for quarks as
\cite{Berger2001}
\eq \label{eq:GPDs}
V_{\la'\la}&=&\frac{1}{2} \int \frac{d \omega}{2\pi}\, e^{ix (P z)}
  \langle  p', \la' |\, \bar{q}(-\half z)\, \slash{n} \, q(\half z)\,
  \,| p, \la \rangle \Big|_{z = \omega n}
\nonumber \\[0.2em]
&=& \sum_{i} \ep'^{*\nu}  V_{\nu\mu}^{(i)} \ep^\mu  H_i^q(x,\xi,t)
\en
where $\ep=\ep(p,\la)$ [or $\ep'=\ep'(p',\la')$] and $\la$ ($\rm{or}~\la')=0,~\pm1$ are the initial (or final) polarization vector and its helicity, respectively. The explicit expressions of $\ep$ and the helicity amplitudes
of the matrix elements were introduced in Ref.~\cite{Berger2001}. The helicity amplitudes give the connection
between GPDs and the Deep Inelastic Scattering(DIS) structure functions by taking the forward limit. It is argued that there are five independent
tensor structures that the tensor $V_{\nu\mu}^{(i)}$ in Eq.~({\ref{eq:GPDs}}) would explicit depend on,
\eq
\label{5tensors}
\{ g_{\nu\mu}, P_\nu n_\mu , n_\nu P_\mu , P_\nu P_\mu , n_\nu n_\mu  \} \ .
\en
Consequently, the GPDs of the $\rho$ meson are defined as
\eq \label{eq:5GPDs}
V_{\la'\la}&=& -(\ep'^*\cdot\ep)H_1^q +\frac{(\ep\cdot n)(\ep'^*\cdot P)+(\ep'^*\cdot n)(\ep\cdot P)}{P\cdot n}H_2^q
-\frac{2(\ep\cdot P)(\ep'^*\cdot P)}{M^2}H_3^q
\nonumber \\ &&
+\frac{(\ep\cdot n)(\ep'^*\cdot P)-(\ep'^*\cdot n)(\ep\cdot P)}{P\cdot n}H_4^q
+\left\{ M^2\frac{(\ep\cdot n)(\ep'^*\cdot n)}{(P\cdot n)^2} + \frac{1}{3}(\ep'^*\ep) \right\} H_5^q \ ,
\en
where $M$ is the $\rho$ meson mass. The five unpolarized GPDs $H_{i}^q$($i=1\sim 5$) are the functions of $x$, $\xi$, and $t$. The explicit dependence of $H_i^q$ on the three variables is omitted for simplicity.\\

{\it Sum rules.}-The conventional form factor decomposition of the vector current for a spin-1 particle is
\eq\label{eq:Imm}
I_{\la'\la}^\mu &=& \langle  p',  {\la'} |\, \bar{q}(0)\, \gamma^\mu \, q(0)\,| p, \la \rangle
\nonumber \\ 
&=& \ep'^{*\beta}  \ep^\alpha  \bigg[ -\Big( G_1^q(t) g_{\beta\alpha}
+ G_3^q(t)  \frac{ P_\beta P_\alpha}{2M^2} \Big) P^\mu
+ G_2^q(t) \left( g_\alpha^\mu P_\beta  + g_\beta^\mu P_\alpha \right)
 \bigg]  \ .
\en
The conventional FFs $G_{1,2,3}$ are obtained from $G_{1,2,3}^q$ by weighting with electromagnetic charges and then summing over flavors: $G_{i}=e_u G_{i}^u + e_{\bar{d}} G_{i}^{\bar{d}}$ for $i=1, 2, 3$. It is equivalent to using the isospin combination, which will be shown later in Eq.~(\ref{eq:Gi}). Comparing with Eq.~(\ref{eq:5GPDs}), one can obtain the sum rules,
\eq\label{eq:sumrule}
\int_{-1}^{1} dx H_i^q (x,\xi,t) &=& G_i^q(t) \quad (i=1,2,3) \ , \nonumber \\
\int_{-1}^{1} dx H_i^q (x,\xi,t) &=& 0  \quad (i=4,5) \ .
\en
The integrals of $H_4^q$ and $H_5^q$ vanish due to the constraints of time reversal and Lorentz invariance, respectively~\cite{Berger2001}.\\

The FFs $G_{C,M,Q}$ can be expressed in terms of $G_{1,2,3}$ as~\cite{Choi2004}
\begin{eqnarray}
G_C(t)&=&G_1(t) + \frac{2}{3}{\eta} G_Q(t) \ , \ \nonumber \\
G_M(t)&=&G_2(t) \ , \ \nonumber \\
G_Q(t)&=&G_1(t) - G_2(t) + (1+\eta)G_3(t)\ , \
\label{eq:Gcmq}
\end{eqnarray}
where $\eta=-t/4M^2$. Together with Eq.~(\ref{eq:sumrule}), one can obtain $G_{C,M,Q}$ directly from GPDs $H_{1,2,3}$.
Note that in many previous studies, the calculation of $G_{C,M,Q}$ from the matrix elements of $I_{\la'\la}^+$
is faced with the well-known ambiguity of the angular condition~\cite{Melo1997}.
Some different prescriptions are proposed to avoid the ``worst" matrix elements.
The present work bypasses this ambiguity.

The normalizations take
\eq
G_C(0)=1 \ ,~~~ G_M(0)=2M \mu \ , ~~~G_Q(0)=M^2 Q_\rho \ ,
\en
where $\mu$ and $Q_\rho$ are the $\rho$ magnetic dipole and quadrupole moments.
The mean square charge radius $<r^2>$ is given by
\eq
<r^2>=\lim_{t\rightarrow 0}\frac{6\left[ G_C(t)-1 \right]}{t} \ .
\en

{\it Forward limit.}---For $x > 0$, the helicity amplitudes  in the forward limit ($\Delta=0$) give the relations
between GPDs and the unpolarized (quark-spin-averaged) parton distributions $q^\lambda(x)$~\cite{Best1997,Berger2001},
with $\lambda$ being the polarization of the $\rho$ meson, as
\eq
H_1^q(x,0,0) &=& \frac{q^1(x)+q^{-1}(x)+q^0(x)}{3} = q(x) \ , \nonumber \\
H_5^q(x,0,0) &=& q^0(x) - \frac{q^1(x)+q^{-1}(x)}{2} \ .
\en
For $x < 0$, the above equations with an overall sign change give the antiquark distributions at $-x$. Here, the
unpolarized quark density is defined as $q^\lambda=q^\lambda_{\uparrow}+q^\lambda_{\downarrow}$,  where $\uparrow$ ($\downarrow$) stands for up (down) spin projection along the direction of the motion when the $\rho$ meson moves with
infinite momentum. In the constituent quark model, the sum rules, corresponding to the flavor number and momentum conservation, are
\eq
\label{eq:sumrule_q0}\int dx u(x) = \int dx \bar{d}(x) &=& 1 \ ,  \\
\label{eq:sumrule_q1}\int dx [ x \left( u(x)+\bar{d}(x) \right) ] &=& 1 
\en
for the $\rho^+$ meson.\\

At leading twist or leading order, the single flavor DIS structure function
$F_1^q (x)$ is one-half of the probability to find a quark with momentum
fraction $x$ and obeys the Callan-Gross relation~\cite{Berger2001,Best1997}
\eq
\label{eq:F1q}
F_1^q (x) = \frac{1}{3} \left[ q^1_\uparrow(x) +q^1_\downarrow(x) + q^0_\uparrow(x) \right]
= \frac{1}{2} H_1^q(x,0,0)
\ .
\en
The single flavor structure function $b_1^q(x)$, which measures the difference in the spin projection of the $\rho$ meson,
only depends on the quark-spin-averaged distribution $q^\lambda(x)$,
\eq
b_1^q(x) &=&  q^0(x) - \frac{q^1(x)+q^{-1}(x)}{2} = H_5^q(x,0,0) \ .
\en
From parity, one has $q^\lambda_{\uparrow}=q^{-\lambda}_{\downarrow}$, and therefore the conventional structure functions, related to $q^\lambda(x)$, are
\eq
F_1(x) = \sum_{q} e_q^2 F_1^q (x)  \ , \quad 
b_1(x) = \frac{1}{2} \sum_{q} e_q^2 b_1^q(x) \ .
\en
In the following, we will focus on the single flavor structure functions. In the meson case, the structure functions are identical for both flavors. It should be mentioned that the spin-1 particle, different from the spin-1/2 one, has
the tensor structure function $b_1$. It triggers great interest~\cite{Close:1990zw,Cosyn:2017fbo,Dong:2014eya, Kumano:2010vz,Khan:1991qk}. The sum rule of this structure function is $\int dx b_1(x)=0$~\cite{Close:1990zw}.\\

In addition, the $n$th Mellin moment of a function $f(x)$ is defined as
\eq
M_n(f)=\int_0^1 x^{n-1}f(x)dx \ .
\en
For the $\rho$ meson case, to the leading order (twist 2), one finds
\eq
2M_n(F_1^q)=C_n^{(1)}a_n \ , \quad
2M_n(b_1^q)=C_n^{(1)}d_n \ ,
\en
where $C_n^{(k)}=1+O(\alpha)$ are the Wilson coefficients of the operator product expansion and $a_n$ and $d_n$
are the reduced matrix elements~\cite{Best1997}. With the quenched approximation, Ref.~\cite{Best1997} found that
these relations hold for both even and odd $n$th orders.\\

{\it Isospin combination.}---In Eq.~(\ref{eq:5GPDs}), GPDs are defined flavor by flavor. Similar to Refs.~\cite{Broniowski2008,Broniowski200878}, the corresponding
isospin projection of the isovector ($I=1$, nonsinglet) equals
\begin{eqnarray}
\lefteqn{
\frac{1}{2} \int \frac{d \omega}{2\pi}\, e^{ix (P z)}
  \langle \rho^b (p',\la')|\, \bar{q}(-\half z)\, \slash{n} \, \tau_3 q(\half z)\,
  \,|\rho^a (p,\la) \rangle \Big|_{z = \omega n}}
\nonumber \\[0.2em] &&
    = \imath \epsilon_{3ab} \Bigg\{
    - (\epsilon'^*   \epsilon)\, H_{1}^{I=1}
    + \frac{(\epsilon n) (\epsilon'^* P)+ (\epsilon'^* n) (\epsilon P)}{P n}\, H_{2}^{I=1}
    - \frac{2 (\epsilon P)(\epsilon'^* P)}{M^2}\, H_{3}^{I=1}
\nonumber \\[0.2em] && \quad
    + \frac{(\epsilon n) (\epsilon'^* P) - (\epsilon'^* n) (\epsilon P)}{P n}\, H_{4}^{I=1}
    + \Bigg[
    M^2\, \frac{(\epsilon n)(\epsilon'^* n)}{(P n)^2}
    +\frac{1}{3} (\epsilon'^*   \epsilon) \Bigg] H_{5}^{I=1} \; \Bigg\}
\nonumber \\[0.2em] && {}=
    \imath \epsilon_{3ab} \Bigg\{ - (\epsilon'^*   \epsilon)\, \left( H_{1}^{u}- H_{1}^{d} \right)
    + \frac{(\epsilon n) (\epsilon'^* P)
    + (\epsilon'^* n) (\epsilon P)}{P n}\, \left( H_{2}^{u}- H_{2}^{d} \right)
    - \frac{2 (\epsilon P)(\epsilon'^* P)}{M^2}\, \left( H_{3}^{u}- H_{3}^{d} \right)
\nonumber \\[0.2em] && {}\quad
    + \frac{(\epsilon n) (\epsilon'^* P)
    - (\epsilon'^* n) (\epsilon P)}{P n}\, \left( H_{4}^{u}- H_{4}^{d} \right)
    + \Bigg[ M^2\, \frac{(\epsilon n)(\epsilon'^* n)}{(P n)^2}
     +\frac{1}{3} (\epsilon'^*   \epsilon) \Bigg] \left( H_{5}^{u}- H_{5}^{d} \right) \; \Bigg\}.
\label{eq:isovectorH}
\end{eqnarray}
where $a,b$=0,1,2, and $\rho^\pm=\rho^1\mp\imath\rho^2$. For the isoscalar case ($I=0$, singlet), one needs the exchange $\slash{n} \, \tau_3 \leftrightarrow \slash{n}$, and therefore $H_{i}^{u}- H_{i}^{d} \leftrightarrow H_{i}^{u}+ H_{i}^{d}$. In the following work, we will only deal with a positive-charged $\rho$ and omit the subscript $+$ whenever no ambiguity arises.\\

\section{Our approach}
\label{sec:The_model}
\begin{figure}
\centering
{\hskip -1.5cm}
\subfigure[The struck $u$ quark in the valence regime]{\label{fig:loopsud:u}
\begin{minipage}[b]{0.4\textwidth}
\includegraphics[width=1\textwidth]{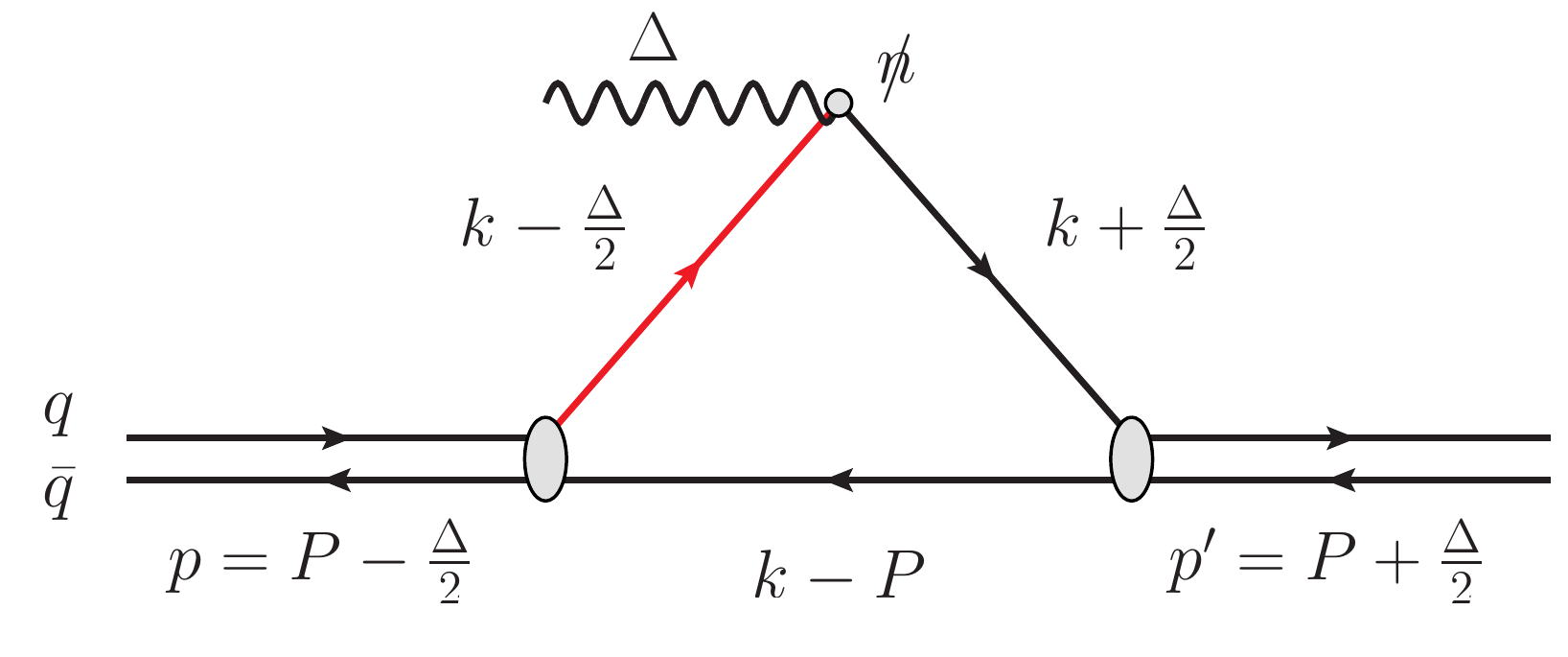} 
\end{minipage}
}
{\hskip 1cm}
\subfigure[The struck $d$ quark in the valence regime]{\label{fig:loopsud:d}
\begin{minipage}[b]{0.3\textwidth}
\includegraphics[width=1\textwidth]{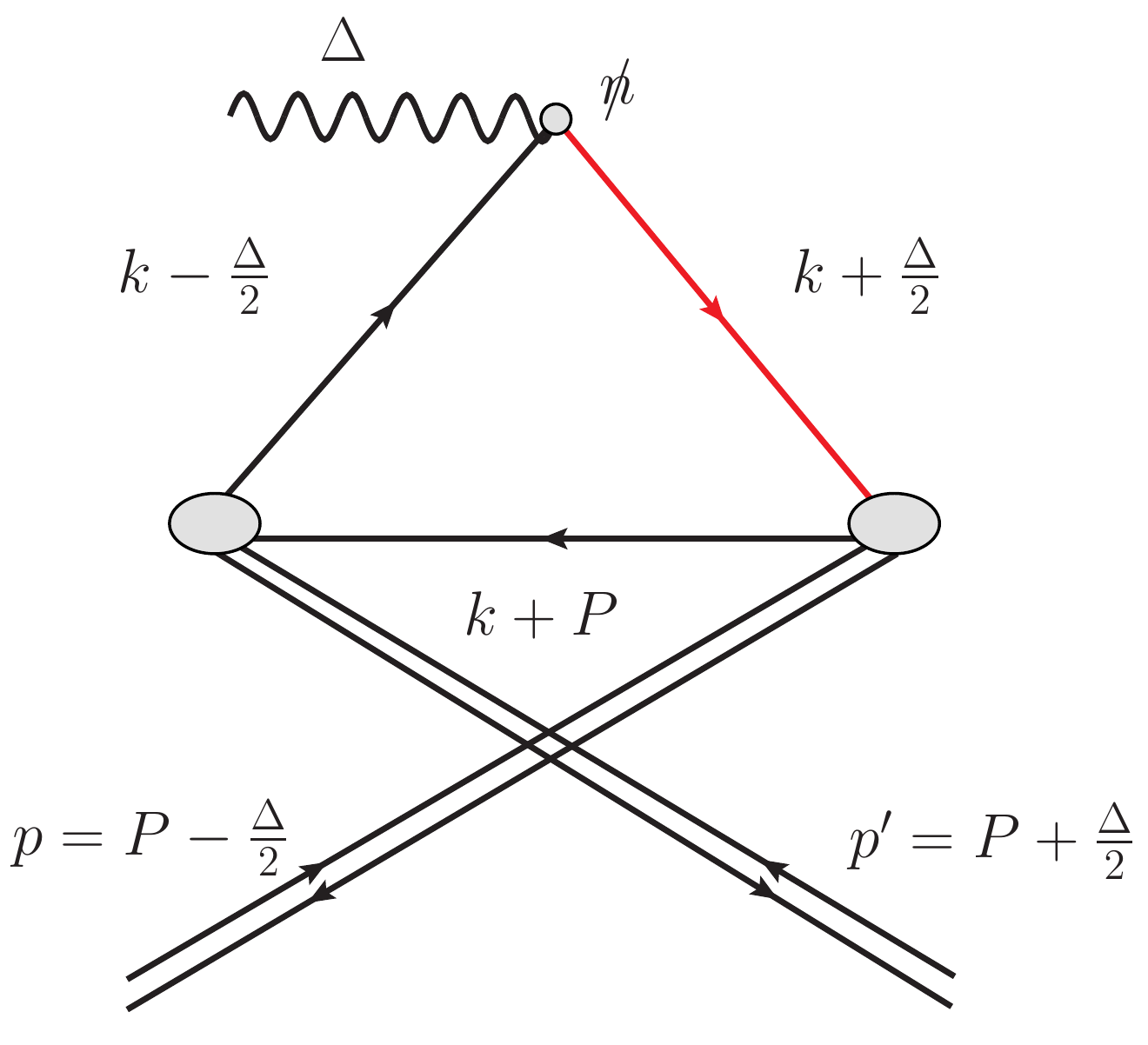} 
\end{minipage}
}
\caption{\label{fig:loopsud}{\small Direct $(a)$ and crossed $(b)$ Feynman diagrams contributing to the GPDs of quark $q$
and antiquark $\bar{q}$ of $\rho$ meson. The gray blobs represent the normal Light Front(LF) wave vertexes. The plus component
of the momentum carried by red lines have positive sign in the valence regime.
}}
\end{figure}

\begin{figure}[t]
\centerline{\includegraphics[width=8.6cm]{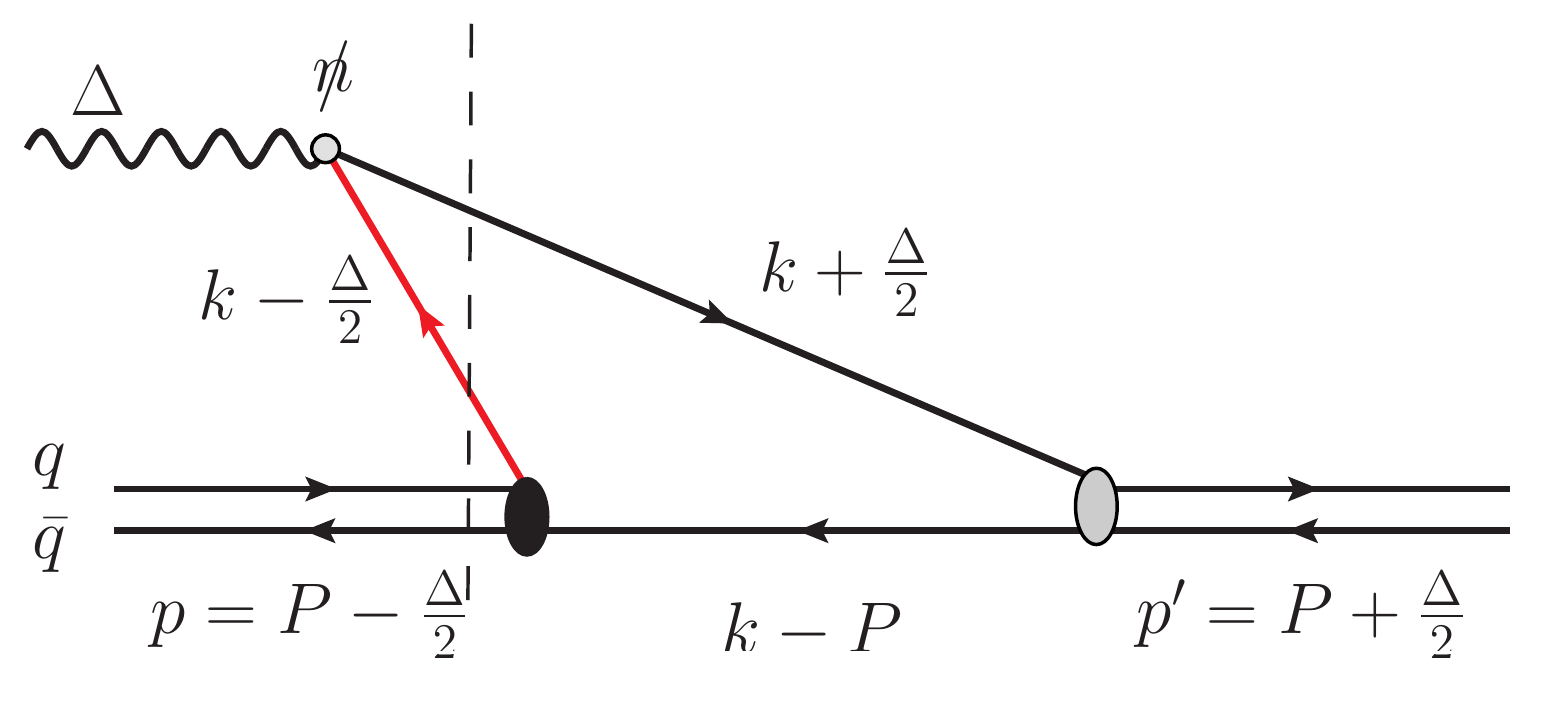}} 
\caption{\label{fig:nonval}
\small
The struck $u$ quark in the nonvalence regime, yielded by the off-diagonal terms in the Fock space. The black blob represents
the non-wave-function vertex. The red line has the negative sign in this regime.
}
\end{figure}

In analogy to the chiral interaction Lagrangian for the $\pi\rightarrow q\bar{q}$ vertex~\cite{Frederico1992},
the effective Lagrangian for the $\rho\rightarrow\bar{q}q$ is taken as
\begin{eqnarray}\label{key}
\mathcal{L}_I &\sim & -\imath \frac{M}{f_\rho} \bar{q} \gamma^\mu \mathbf{\tau} q \cdot \mathbf{\rho}_\mu
\nonumber \\ &&
= -\imath \frac{M}{f_\rho}
\Big[ \bar{u}  \gamma^\mu u  \rho^0_\mu
+ \sqrt{2} \bar{u}  \gamma^\mu d  \rho^+_\mu
+ \sqrt{2} \bar{d}  \gamma^\mu u  \rho^-_\mu
+ \bar{d}  \gamma^\mu d  \rho^0_\mu
\Big] \ ,
\end{eqnarray}
where $f_\rho$ is the $\rho$ decay constant. In the lowest Fock state, the two-parton correlation function, the lhs of Eq.~(\ref{eq:isovectorH}), corresponds to a triangle loop~\cite{Ji2006}. The loop integral, corresponding
to the active $u$ quark [see Fig.~\ref{fig:loopsud:u} and \ref{fig:nonval}], is specified as
\begin{eqnarray}\label{eq:u_quark}
V^{u}_{} (x, \xi, t) &=&
  N_{\mu\nu} \int \frac{d^4 k  }{ (2\pi)^4 }
  \, \delta \left[ n \cdot ( x P - k ) \right]
  (-) Tr \Bigg[
  \frac{\imath ( \slash{k}-\slash{P}+m ) }{ (k-P)^2-m^2 + \imath \epsilon}
  \gamma^\nu
  \frac{\imath ( \slash{k}+\frac{\slash{\Delta}}{2} +m ) }{ (k+\frac{\Delta}{2})^2 -m^2 + \imath \epsilon}
  \slash{n}
    \nonumber \\  && \times
  \frac{\imath ( \slash{k}-\frac{\slash{\Delta}}{2} +m ) }{ (k-\frac{\Delta}{2})^2 -m^2 + \imath \epsilon}
  \gamma^\mu
  \Bigg]
  \Lambda(k-P,p')~
  \Lambda(k-P,p),
\end{eqnarray}
where $m$ is the constituent quark mass and
\eq
N_{\mu\nu} &=& \frac{M^2}{f_\rho^2}
  \frac{  \epsilon'^*_\nu(p', \la') \epsilon_\mu(p, \la) }{ 2(2\pi)^3 \sqrt{\omega_{p'}\omega_{p} } } \ ,
\en
and the scalar function
\eq
\label{eq:lambda}
\Lambda(k-P,p) &=& \frac{c }{ [ (k-P)^2-m^2_R+ \imath \epsilon] [ ({k}-\frac{\Delta}{2})^2-m^2_R+ \imath \epsilon] }
\en
is following Ref.~\cite{Frederico2009}, with $m_R$ and $c$ being the regulator mass and the normalization factor, respectively. The loop
of the struck $d$ quark can be obtained from the crossed Feynman diagram of Fig.~\ref{fig:loopsud:d}. Here, the
scalar product function $\Lambda(k-P,p)$ is symmetric under the exchange of the momentums of the two constituents.
This scalar function is employed to describe the momentum dependent between $q$ and $\bar{q}$ inside the $\rho$
meson. Actually, it plays a role of the momentum cutoff similar to the Pauli-Villars  regularization~\cite{Frederico2009}. It may also stand for a property of the Bethe-Salpeter amplitude~\cite{Melo2002} and contain the information of the nonperturbative effect. Conceptually, by taking $\Lambda(k-P,p)$ as a part of the quark-antiquark-meson vertex, one gets the smeared quark-antiquark-meson vertex, $\gamma^\mu\Lambda(k-P,p)$~\cite{Choi2004}. As will be seen later, the symmetric momenta convention, shown in Figs.~\ref{fig:loopsud} and \ref{fig:nonval}, enables the vertex to fulfill the constraint from the isospin symmetry.\\

Note that the elastic FFs can be calculated in different reference frames, such as the Drell-Yan frame~\cite{Brodsky1998}, where $\Delta^+=0$ and $\xi=0$, and the Breit frame, where $\Delta^+=-\Delta^-$ (see Ref.\cite{Frederico2009} and \cite{lev1998} for discussions on the motivation of adopting this frame). In this work, the above loop integral is performed in the Breit frame, and then the $\xi$ dependence of GPDs can be obtained as well. In this special reference frame, the momentum transfer and initial and finial momenta are
\eq
\Delta &=& (\Delta^+,\Delta^-,\bm\Delta_\perp)=(q_z,-q_z,q_x,q_y) \ , \nonumber \\
p &=& (p^0- \frac{q_z}{2},p^0+\frac{q_z}{2},-\frac{\bm\Delta_\perp}{2})  \ , \nonumber \\
p' &=& (p^0+\frac{q_z}{2},p^0-\frac{q_z}{2},\frac{\bm\Delta_\perp}{2})  \ , 
\en
where $\bm\Delta_\perp=(q_x,q_y)$ and $p^0=M\sqrt{1-\Delta^2/4M^2}$.
Since $\bm\Delta_\perp^2\geqslant 0$, one gets the constraint $\abs{\xi}\leqslant 1/\sqrt{1-4M^2/t}$. \\

The physics in the nonvalence regime, shown in Fig.~\ref{fig:nonval}, is remarkably different from the one in the valence regime. According to Ref.~\cite{Burkardt2000}, the $q\bar{q}$ pair, created by the virtual photon, could interact with itself and form a virtual meson, before merging with the meson state. From another point of view, the higher Fock component contributions should be taken into account in both two regimes for completeness. Instead of finding all higher Fock component contributions as Refs.~\cite{Brodsky2001,Diehl2001}, we handle the nonvalence contribution by replacing the simple $\gamma^\mu$ with a phenomenological meson vertex $\Gamma^\mu$ as shown in Ref.~\cite{Choi2004}. This is an analogy to the covariant form~\cite{Jaus2003}, which has been applied for the deuteron case in our previous work~\cite{Sun2016}.
Thus the smeared quark-antiquark-meson vertex becomes $\Gamma^\mu\Lambda(k-P,p)$.\\

For the $u$ quark contribution shown in Fig.~\ref{fig:loopsud:u}, the spectator constituent momentum is $k_s=k_{\bar{d}}=k-P$. The phenomenological vertices under this loop momentum assignment read
\begin{eqnarray}
\label{eq:vertex}
\Gamma^{\mu}_{i} =
    \gamma^{\mu} - \frac{(2k-P-\frac{\Delta}{2})^{\mu}}{D_{i}} \ , \quad
\Gamma^{\nu}_{f} =
    \gamma^{\nu} - \frac{(2k-P+\frac{\Delta}{2})^{\nu}}{D_{f}} \ ,
\end{eqnarray}
where $D_{i,f} = M_{i,f} + 2m $, and the kinematic invariant masses $M_{i,f}$ are~\cite{Choi2004}
\begin{eqnarray}
\label{eq:vertexM:v}
M_{i}^2 = \frac{\kappa^2_\perp + m^2}{1-x'} + \frac{\kappa^2_\perp + m^2}{x'} \ , \\
M_{f}^2 = \frac{\kappa'^2_\perp + m^2}{1-x''} + \frac{\kappa'^2_\perp + m^2}{x''} \ ,
\end{eqnarray}
with the LF momentum fractions $x'= - k_s^+/p^+ =(1-x)/(1-\abs{\xi})$, $x''=x' p^+/p'^+ = (1-x)/(1+\abs{\xi}) $,
and
\begin{eqnarray}
\kappa_\perp &=& k_{s\perp} - \frac{k_s^+}{p^+} p_{i\perp} \, =
(k-P)_\perp- \frac{x'}{2} {\Delta}_{\perp} \ , \nonumber \\
\kappa'_\perp &=&
(k-P)_\perp+ \frac{x''}{2} {\Delta}_{\perp} \ . 
\end{eqnarray}
In the nonvalence regime, shown in Fig.~\ref{fig:nonval}, the relation of $-\abs{\xi}<x<\abs{\xi}$ leads to $x' > 1$, and the initial vertex becomes the non-wave-function vertex.
To keep the mass square positive [see $M_i$ in Eq.~(\ref{eq:vertexM:v})], Ref.~\cite{Choi2004} proposes to directly
replace $1-x'$ with $x'-1$ in Eq.~(\ref{eq:vertexM:v}) and gets
\begin{eqnarray}
\label{eq:vertexM:nv}
M_{i(NV)}^2 = \frac{\kappa^2_\perp + m^2}{x'-1} + \frac{\kappa^2_\perp + m^2}{x'}.
\end{eqnarray}
Hereafter, we use the subscripts $V$ and $NV$ to stand for the valence and nonvalence regimes, respectively. Note that, when both the struck and spectator constituents are on mass shells, namely, $(k-\frac{\Delta}{2})^2=(k-P)^2=m^2$, one gets $M_{i}^2=M_{f}^2=M^2$. Due to the exchange $1-x'\leftrightarrow x'-1$, the relation of $M_{i(NV)}^2=M^2$ no longer holds for the nonvalence case. However, $M_{i}^2$ and $M_{i(NV)}^2$ have the same limiting value as $x\rightarrow\abs\xi$, and thus the continuity of GPDs is guaranteed. The physics in the parton-number-changing nonvalence Fock state contributions is much more complicated than that in the valence one, since the creation of the $q\bar{q}$ pair involves an infinite sum of the meson contribution. Due to the lack of the information about the nonvalence regime~\cite{Burkardt2000}, in some model calculations, the discontinuity may arise at $x=\xi$ (or $\abs\xi$) where the valence and nonvalence regimes divide, like in Ref.~\cite{Choi2001} for the $\pi$ meson GPDs.\\

With the above preparations, the integral of Eq.~(\ref{eq:u_quark}) in the light-front frame reads
\begin{eqnarray}\label{Vlf}
V^u_{} (x, \xi, t) &=&
  N_{\mu\nu} \int \frac{dk^+dk^- d{\bm k}_\perp}{ 4 (2\pi)^4}
    \delta\left[ x P^+ -k^+ \right]
    \frac{ (-) Tr[{\cal O}^{\mu\nu+}] }{ (k^+ - P^+) (k^+ +\frac{\Delta^+}{2}) (k^+ -\frac{\Delta^+}{2})}
        \nonumber \\ && ~\times 
    \frac{1 }{ \left[ k^- -P^- -(k-P)^-_{on}+i\frac{\epsilon}{ k^+ - P^+} \right ]}
    \frac{1 }{ \left[ k^- +\frac{\Delta^-}{2} -(k +\frac{\Delta}{2})^-_{on}+i \frac{\epsilon}{k^+ +\frac{\Delta^+}{2}} \right ]}
        \nonumber \\ && ~\times
    \frac{1 }{ \left[ k^- -\frac{\Delta^-}{2} -(k -\frac{\Delta}{2})^-_{on}+i \frac{\epsilon}{k^+ -\frac{\Delta^+}{2}} \right ]}
     \nonumber \\ && ~\times 
     \Lambda(k-P,p')~
    \Lambda(k-P,p) ,
\end{eqnarray}
where
\begin{eqnarray}
{\cal O}^{\mu\nu+} &=& \imath^3
    ( \slash{k}-\slash{P}+m )
    \Gamma^\nu_{i}
    ( \slash{k}+\frac{\slash{\Delta}}{2} +m )
    \gamma^+
    ( \slash{k}-\frac{\slash{\Delta}}{2} +m )
    \Gamma^\mu_{f} \ ,
   \\
\label{eq:kon}
(k-P)^-_{on} &=& \frac{(k-P)_\perp + m_{}^2}{k^+ - P^+} \quad \ , \text{etc.}  \ ,
\end{eqnarray}
and the $\Lambda$ functions, Eq.~(\ref{eq:lambda}), are chosen without changing the distribution of the poles with respect
to the three denominators of the propagators. There are six poles with respect to $k^-$ for the integral, the same as in the pion case~\cite{Frederico2009}. They are
\begin{eqnarray}
k^-_{1(2)} &=& P^- + (k-P)^-_{on(R)} - i\frac{\epsilon}{ k^+ - P^+} \ , \nonumber \\
k^-_{3(4)} &=& \frac{\Delta^-}{2} + (k -\frac{\Delta}{2})^-_{on(R)} - i \frac{\epsilon}{k^+ -\frac{\Delta^+}{2}} \ , \nonumber \\
k^-_{5(6)} &=& -\frac{\Delta^-}{2} + (k +\frac{\Delta}{2})^-_{on(R)} - i \frac{\epsilon}{k^+ +\frac{\Delta^+}{2}} \ ,
\end{eqnarray}
where $(k-P)^-_{R}$ and $(k \pm \frac{\Delta}{2})^-_{R}$ are obtained by replacing $m$ with $m_R$ in Eq.~(\ref{eq:kon}), respectively.
Since $p'^+ > p^+ > 0$, after integrating over $k^-$, there are only two regions in $k^+$ that contribute to the integral, the valence regime  $k^+ \in \left[ {\Delta^+}/{2}, P^+ \right]$ and the nonvalence one $k^+ \in \left[ - {\Delta^+}/{2}, {\Delta^+}/{2} \right]$. In the case $p'^+ = p^+$($\xi=0$), there is only one regime (valence), and the detail calculation can be found in Refs.~\cite{Frederico1992,Miller2009}. The first two poles $k^-=k^-_{1(2)}$ contribute to the valence part, and last two poles $k^-=k^-_{5(6)}$ contribute to the nonvalence one (see Fig. 2). The residue of the pole $k^-=k^-_{1}$ reads
\begin{eqnarray}\label{Vlfv1}
\lefteqn{
V^u_{1(V)} (x, \xi, t) =
   \frac{-N_{\mu\nu} }{ 4 (2\pi)^4} \int_{\frac{\Delta^+}{2}}^{P^+} dk^+ \int d{\bm k}_\perp
    \delta\left[ x P^+ -k^+\right]
    } \ \ \ \ \ \ \ \ \ \ \ \ \ \ \
        \nonumber \\ && ~\times
    \frac{ Tr[{\cal O}^{\mu\nu+}] }{ (k^+ - P^+) (k^+ +\frac{\Delta^+}{2}) (k^+ -\frac{\Delta^+}{2})}
    \frac{1 }{ \left[ k^- +\frac{\Delta^-}{2} -(k +\frac{\Delta}{2})^-_{on}+i \frac{\epsilon}{k^+ +\frac{\Delta^+}{2}} \right ]}
        \nonumber \\ && ~\times
    \frac{1 }{ \left[ k^- -\frac{\Delta^-}{2} -(k -\frac{\Delta}{2})^-_{on}+i \frac{\epsilon}{k^+ -\frac{\Delta^+}{2}} \right ]}
     \nonumber \\ && ~\times 
     \Lambda(k-P,p')~
    \Lambda(k-P,p)
    \Big|_{k^- = k^-_{1}} \ ,
\end{eqnarray}
and for the rest of the poles $k^-=k^-_{i}$, $V^u_{i(V/NV)} (x, \xi, t)$ ($i$ labels different poles)
can be obtained similarly. Then, the valence contributions read
\eq
V^u_{(V)} = V^u_{1(V)} + V^u_{2(V)}  \ ,
\en
where the $\xi$-independent (also frame-independent) full result for the $u$ quark GPDs is
\eq
V^u_{} = V^u_{(V)} + V^u_{(NV)}  \ .
\en
It is easy to verify that, under the assignment of loop momenta in Fig.~\ref{fig:loopsud},
the trace part of the loop integral for the $d$ quark is related to that of the $u$ quark as
\eq
Tr\left[ {\cal O}^{\mu\nu+}_{(d)} (x,-k) \right] = - Tr\left[ {\cal O}^{\mu\nu+}_{} (-x,k) \right] \ .
\en
Therefore, the relation $ V^d_{} ( x, \xi, t) = - V^u_{} ( -x, \xi, t)$ is preserved,
as required by the isospin and crossing symmetries~\cite{Frederico2009}. Here, the symmetric momenta convention are essential for the present model to fulfill this constraint. We, thus, get
\eq
\int_{-1}^1 dx H_i^d (x,\xi,t) =
    -\int_{-1}^{1} dx \; H_i^u (x,\xi,t) \ ,
\en
with $i=1\sim5$. Finally, the isovector GPDs satisfy
    \eq
H_i^{I=1} (x,\xi,t)
    &=& H_i^{I=1} (-x,\xi,t).
\en
In addition, from Eq.~(\ref{eq:sumrule}), the sum rules of the conventional FFs, it is easy to see the equivalent~\cite{Berger2001,Broniowski200878}
\eq \label{eq:Gi}
G_i &=& e_u \int_{-1}^1 dx \; H_i^{u} (x,\xi,t) + e_d \int_{-1}^1 dx \; H_i^d (x,\xi,t)  \nonumber \\
    &=& \int_{-1}^1 dx \; H_i^{I=1} (x,\xi,t) \ .
\en
In our work, the strategy to extract the five GPDs $H_i$, is to construct five independent equations
by replacing $\ep'^{*\nu} \ep^\mu$ in $V^u$ with the tensors listed in (\ref{5tensors}) separately,
and finally to solve them jointly. See the Appendix for more details.

\section{Results}\label{sec:Results}

In this work, we take the constituent mass $m=0.403~\gev$ and regulator mass $m_R=1.61~\gev$. The
requirement of stability of the bound states, $m > M/2$ and $m+m_R > M$, is maintained.\\

\begin{figure}[t]
\centerline{\includegraphics[width=8.6cm]{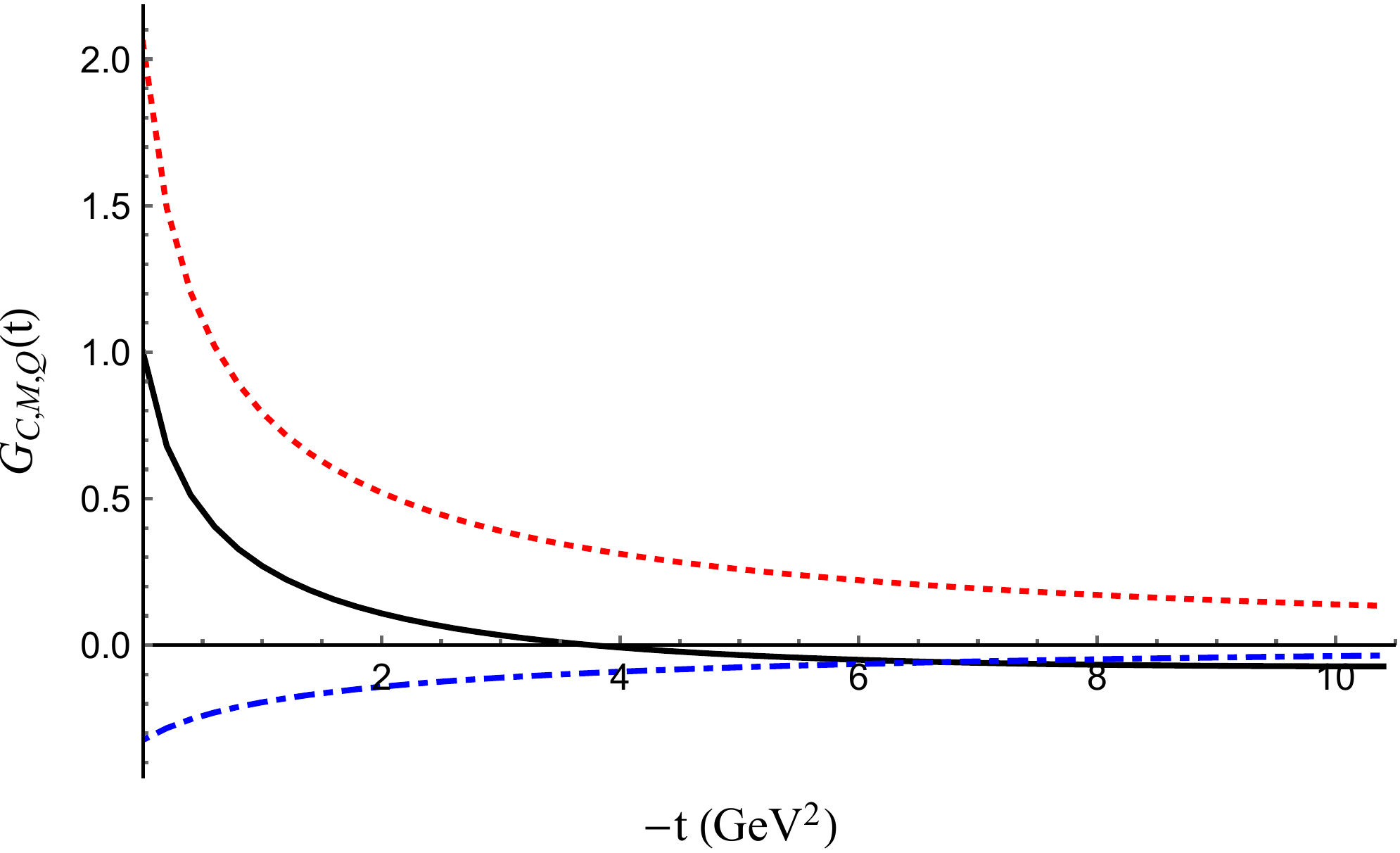}}
\caption{\label{FFs}
The $\rho$ FFs, $G_C$ (solid black line), $G_M$ (dashed red line) and $G_Q$ (dot-dashed blue line).
}
\end{figure}

\begin{figure}[!t]
\centering
\subfigure[]{\label{fig:XiTGV:0}
\begin{minipage}[b]{0.4\textwidth}
\includegraphics[width=1\textwidth]{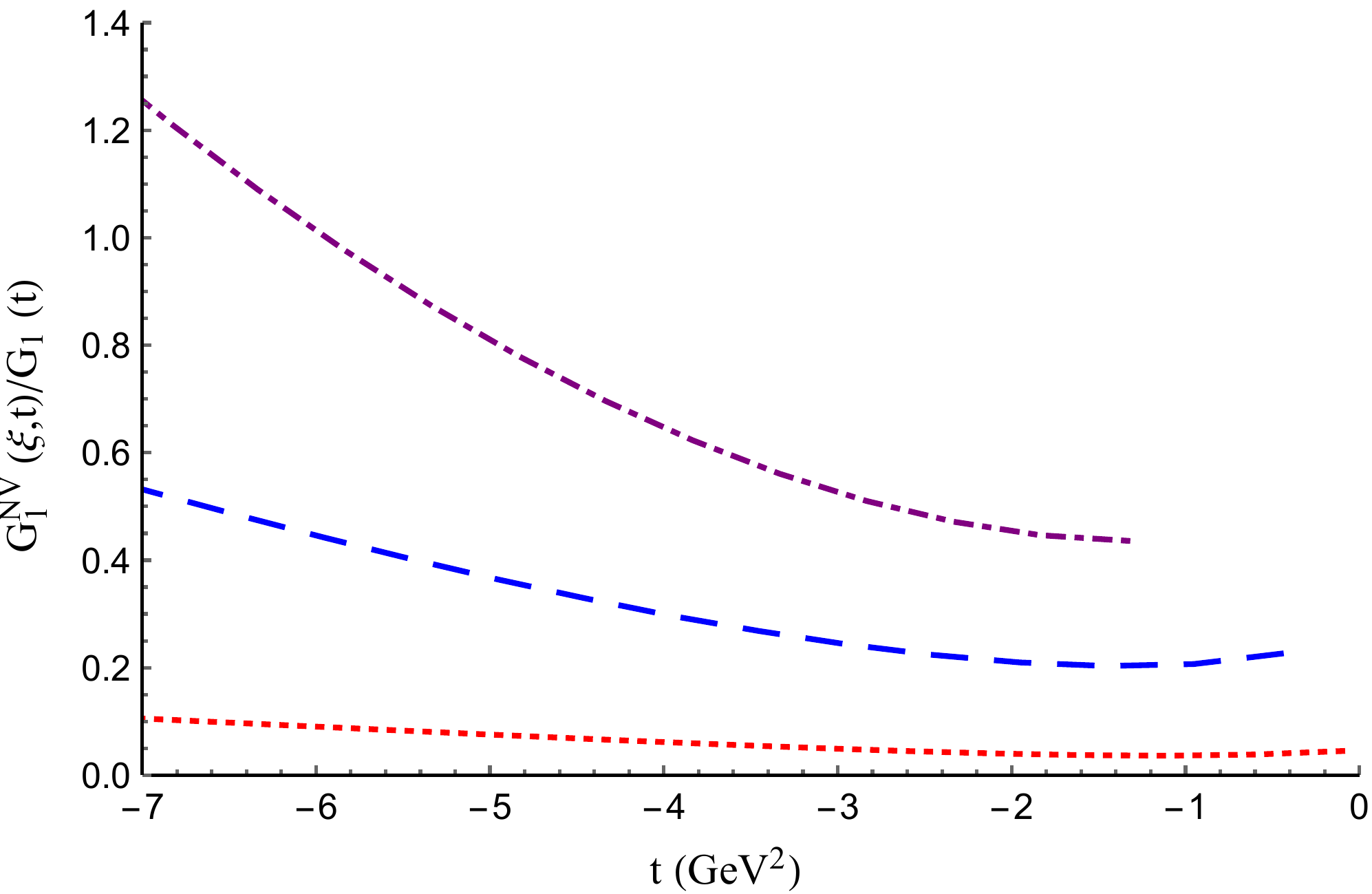} 
\end{minipage}
}
\subfigure[]{\label{fig:XiTGV:1}
\begin{minipage}[b]{0.4\textwidth}
\includegraphics[width=1\textwidth]{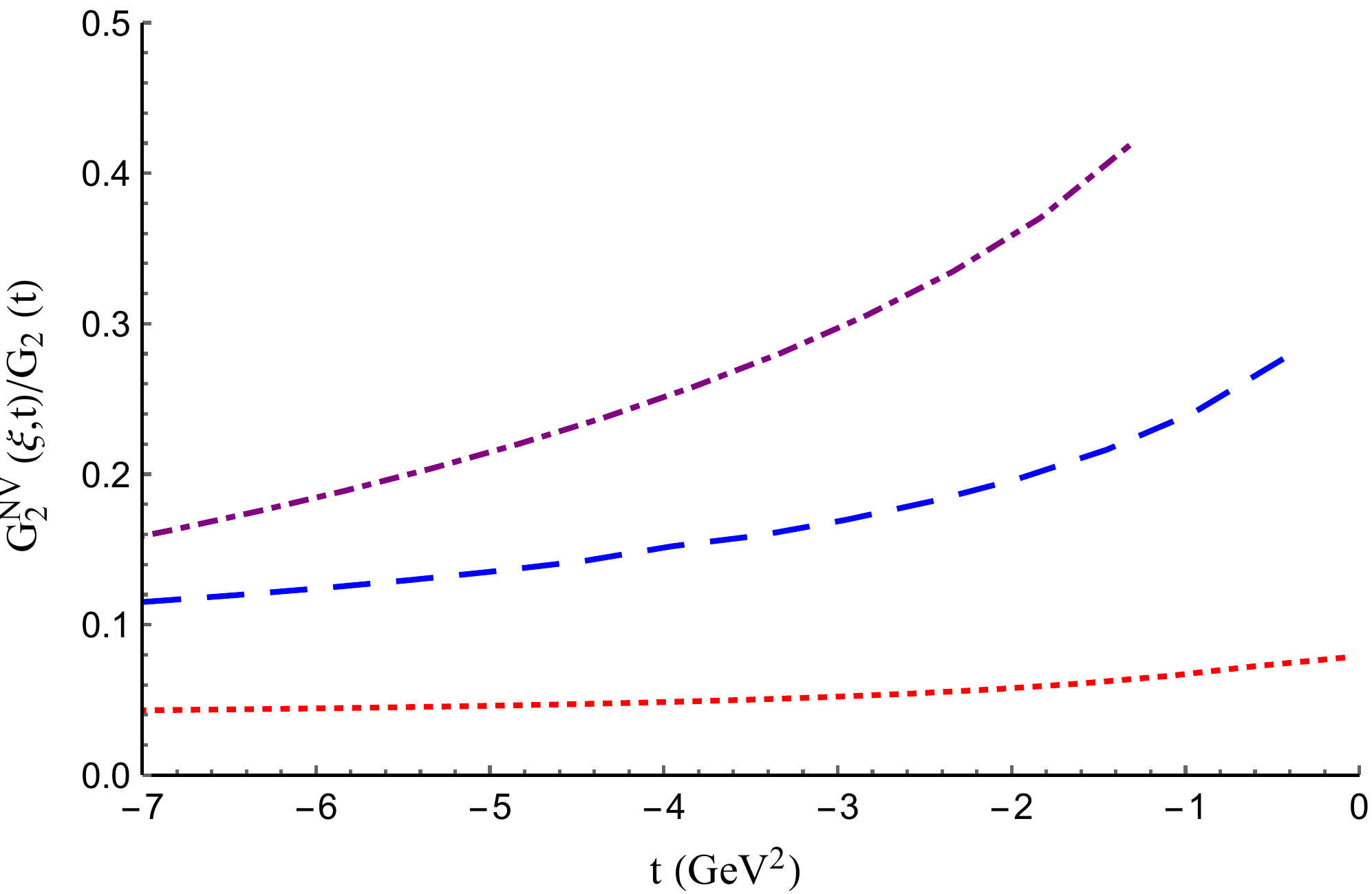} 
\end{minipage}
}
\subfigure[]{\label{fig:XiTGV:2}
\begin{minipage}[b]{0.4\textwidth}
\includegraphics[width=1\textwidth]{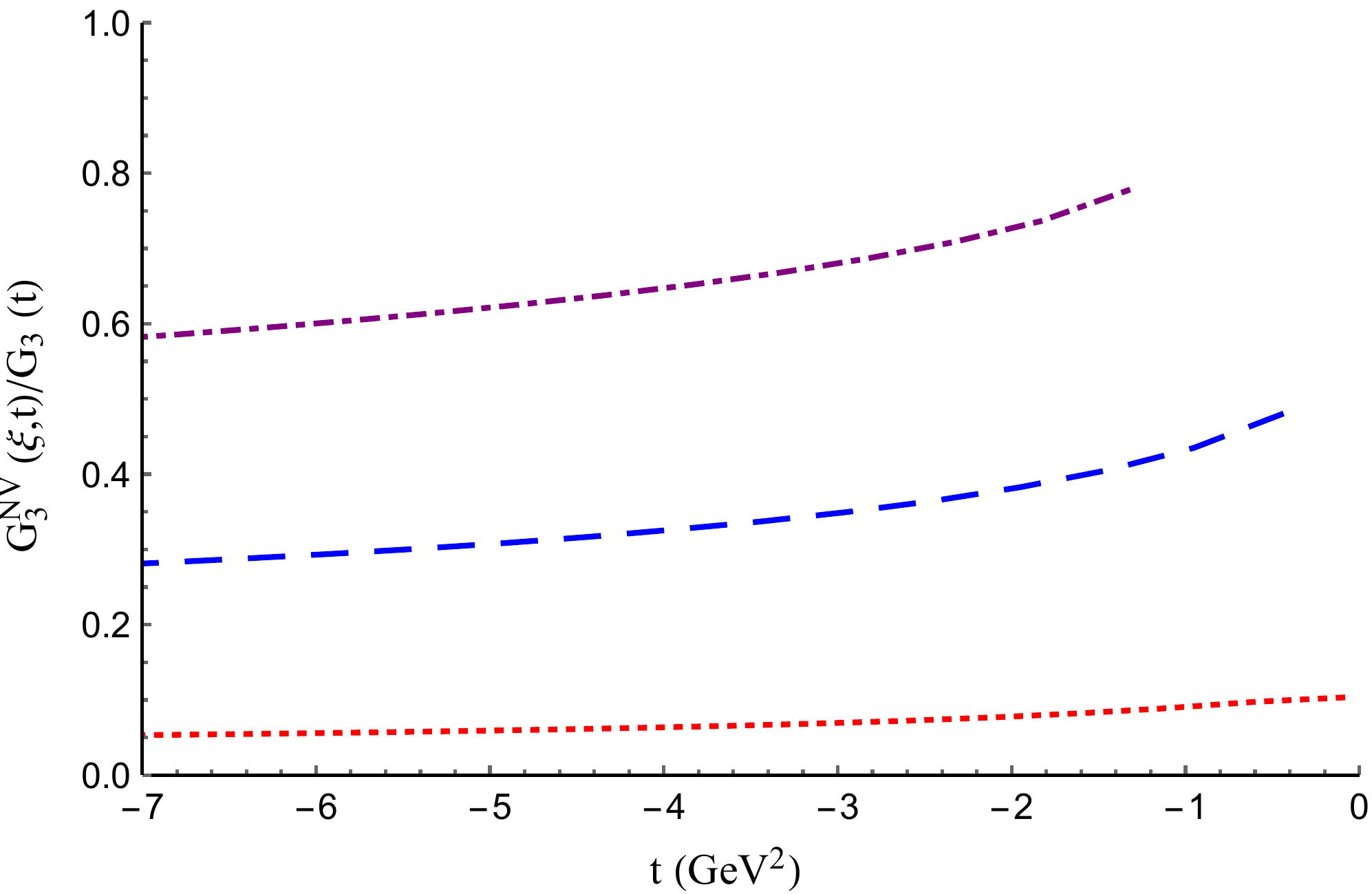}
\end{minipage}
}
\caption{\label{fig:XiTGV}{The nonvalence contributions to FFs $G_1$, $G_2$ and $G_3$ at $\xi$=$-0.2$
\text{(dotted red line)},$-0.4$ \text{(dashed blue line)},$-0.6$ \text{(dot-dashed purple line)}, respectively.}}
\end{figure}

\begin{figure*}[!ht]
\centering
\subfigure[$ \xi=0 $]{\label{fig:h1:3d:xi0}\includegraphics[width=7.0cm]{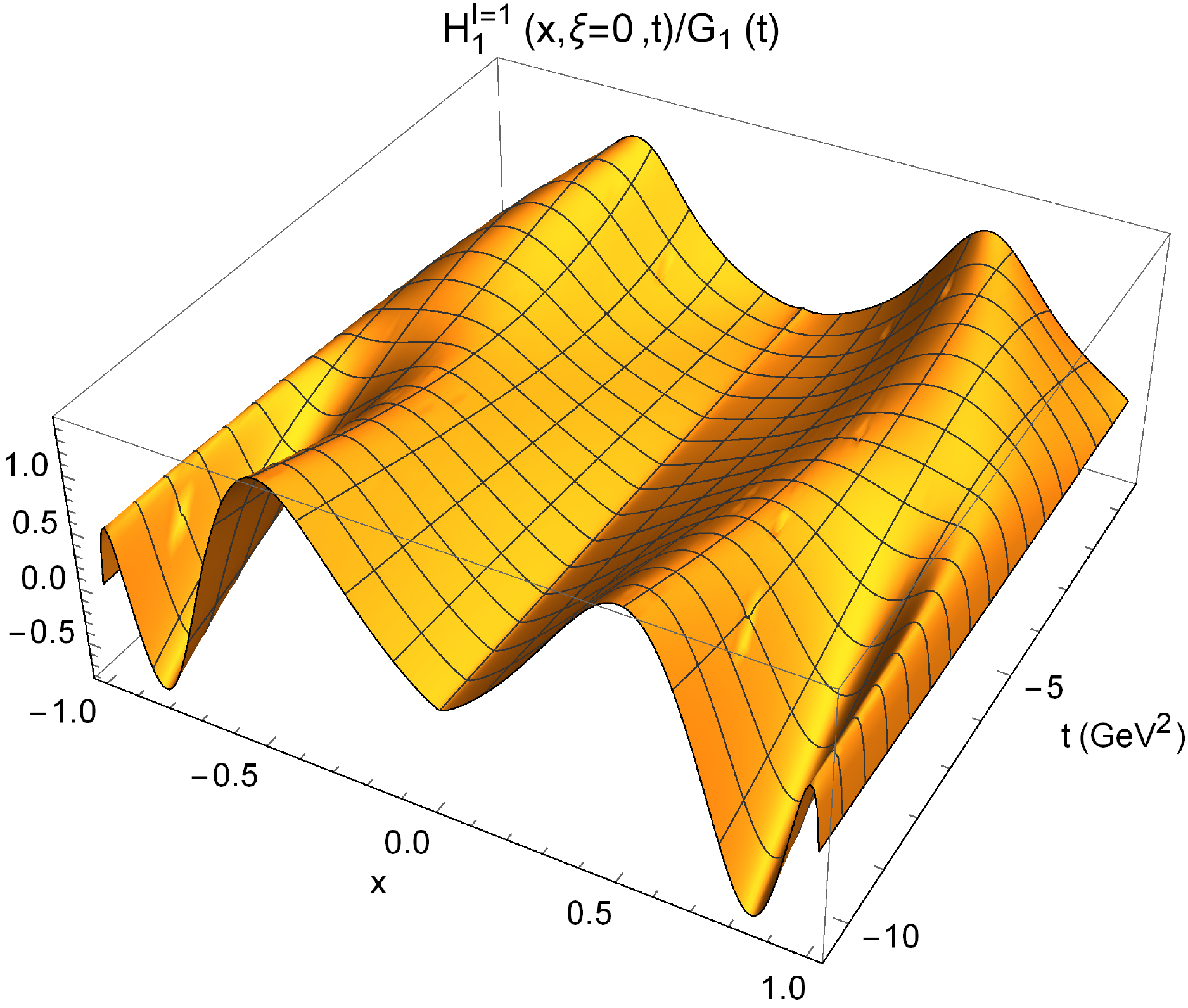}}
\subfigure[$ \xi=-0.4 $]{\label{fig:h1:3d:xi04}\includegraphics[width=7.0cm]{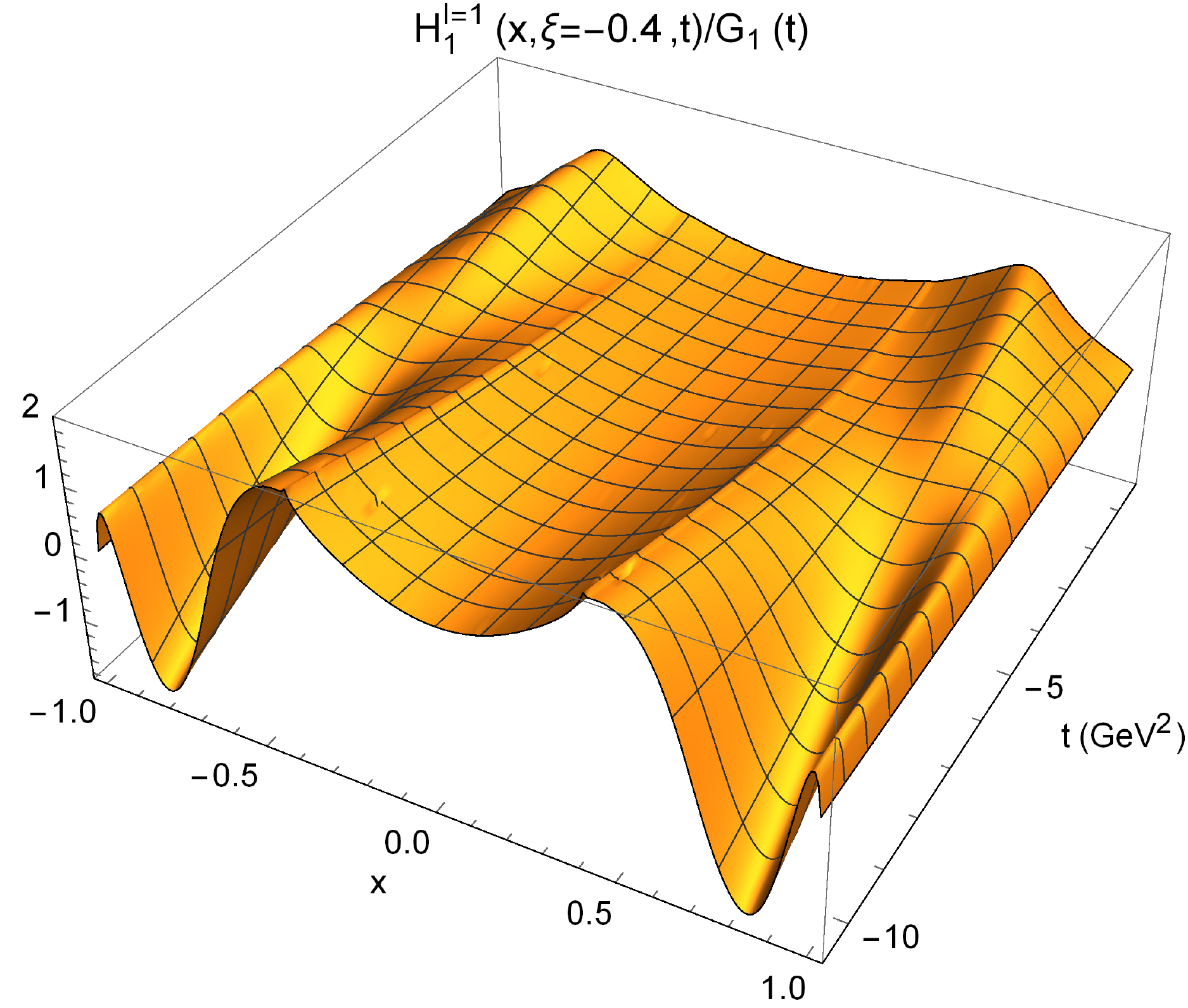}}
\caption{ The 3D $\rho^+$ GPD $H_1$ at $\xi=0~\text{(a)}$ and $-0.4~\text{(b)}$. } \label{fig:h1:3d}
\end{figure*}

\begin{figure*}
\centering
\subfigure[$ \xi=0 $]{\label{fig:h2:3d:xi0}\includegraphics[width=7.0cm]{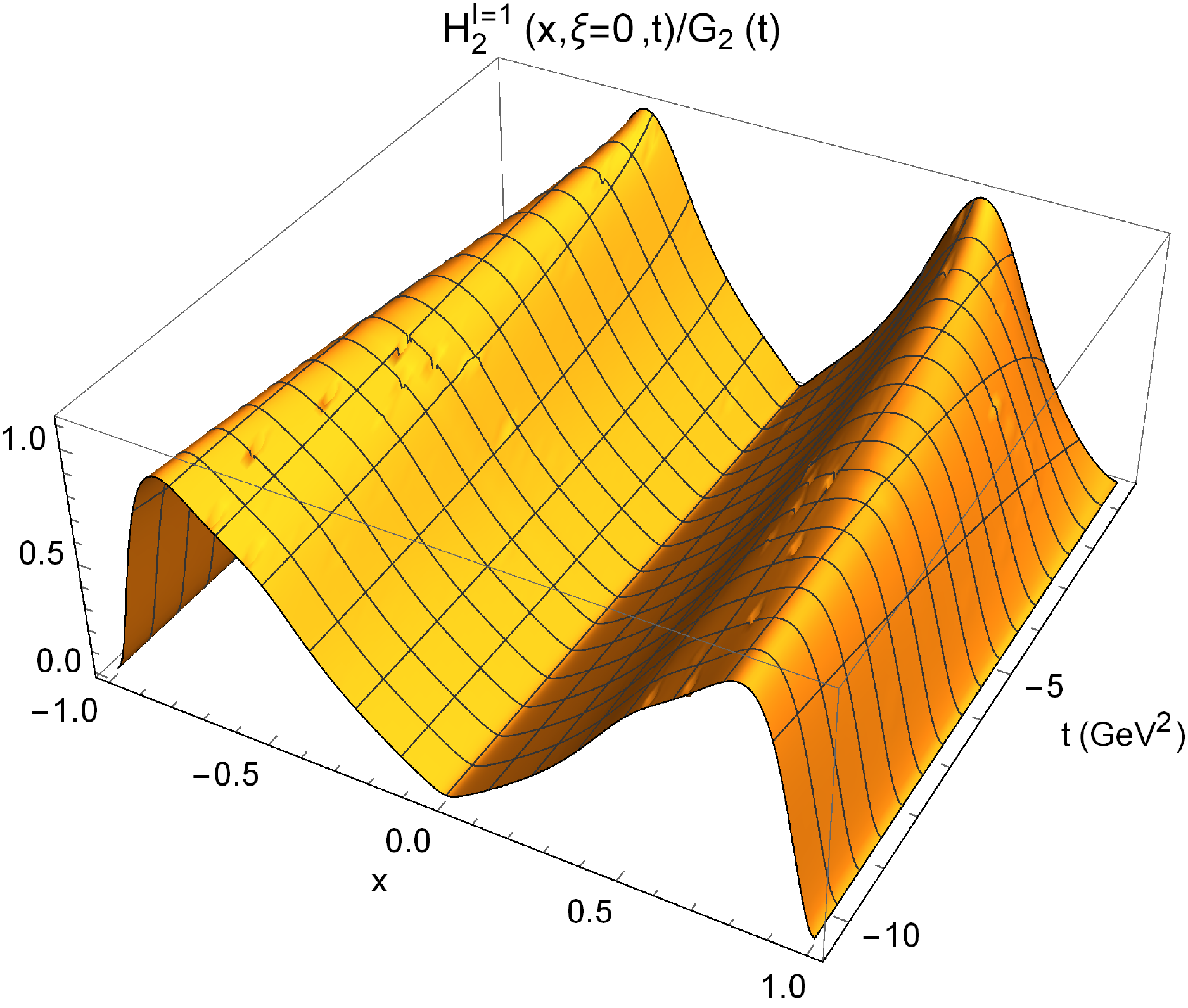}}
\subfigure[$ \xi=-0.4 $]{\label{fig:h2:3d:xi04}\includegraphics[width=7.0cm]{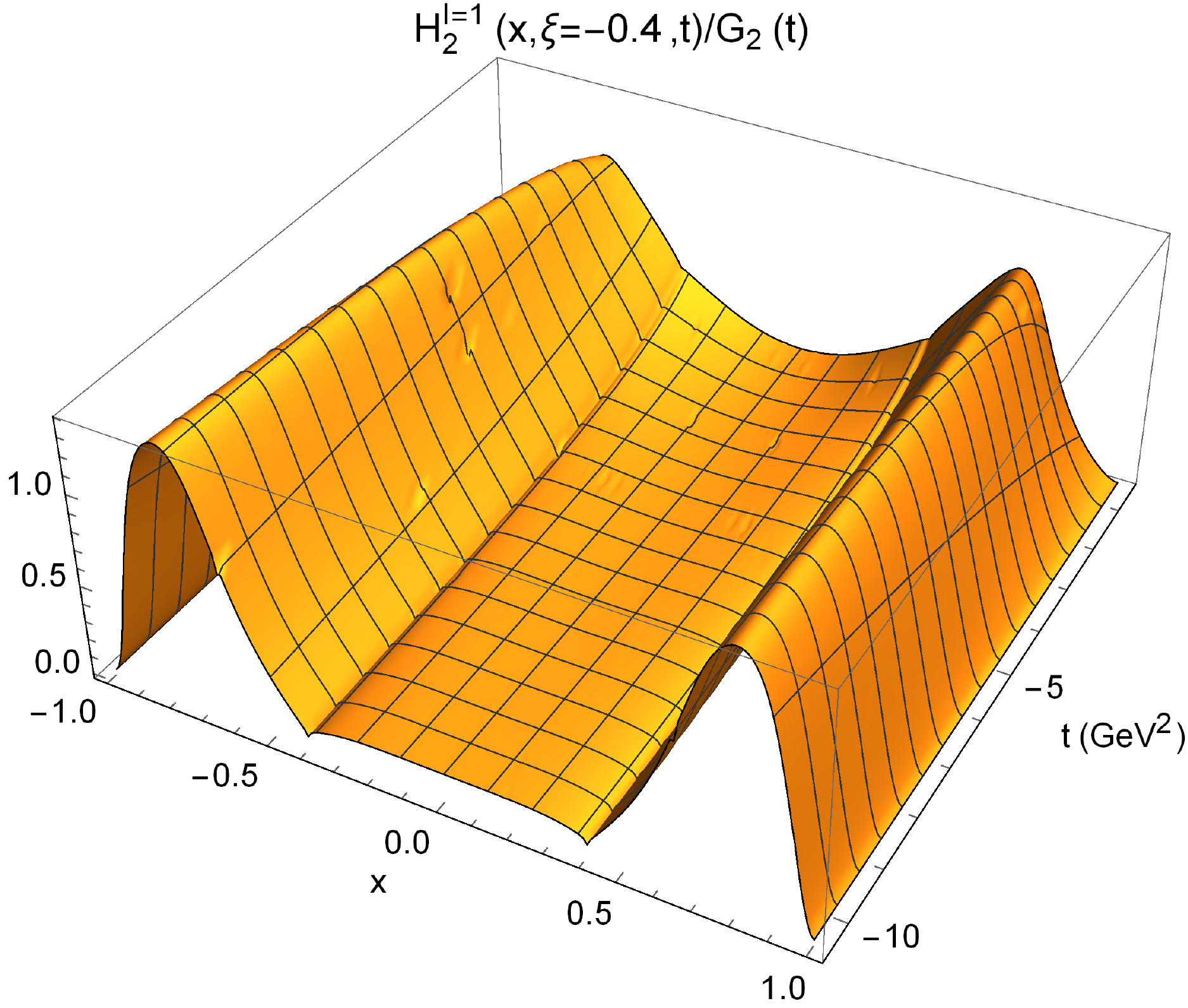}}
\caption{$\rho^+$ GPD $H_2$. The same line code is used in Fig.~\ref{fig:h1:3d}. } \label{fig:h2:3d}
\end{figure*}

\begin{figure*}
\centering
\subfigure[$ \xi=0 $]{\label{fig:h3:3d:xi0}\includegraphics[width=7.0cm]{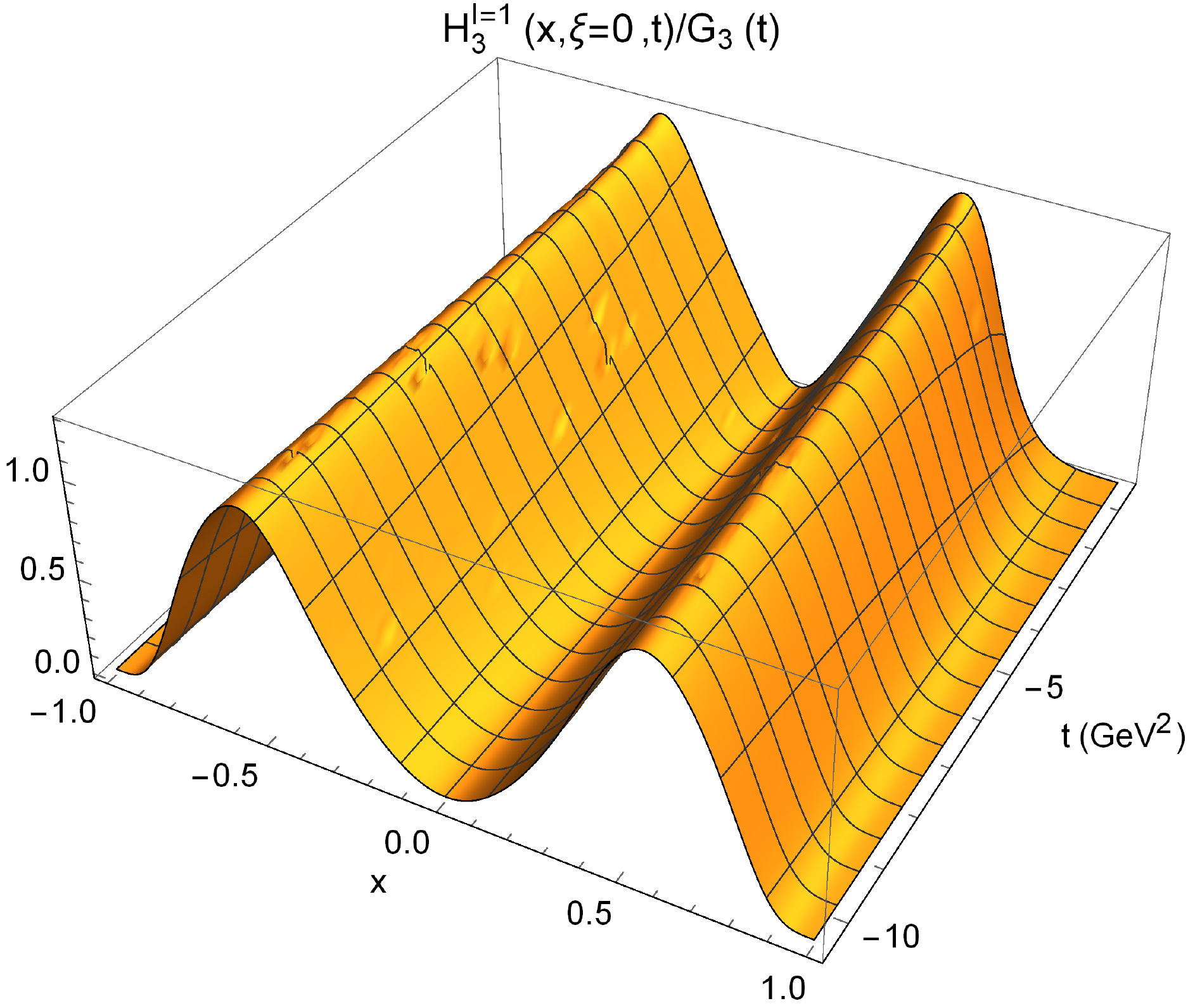}}
\subfigure[$ \xi=-0.4 $]{\label{fig:h3:3d:xi04}\includegraphics[width=7.0cm]{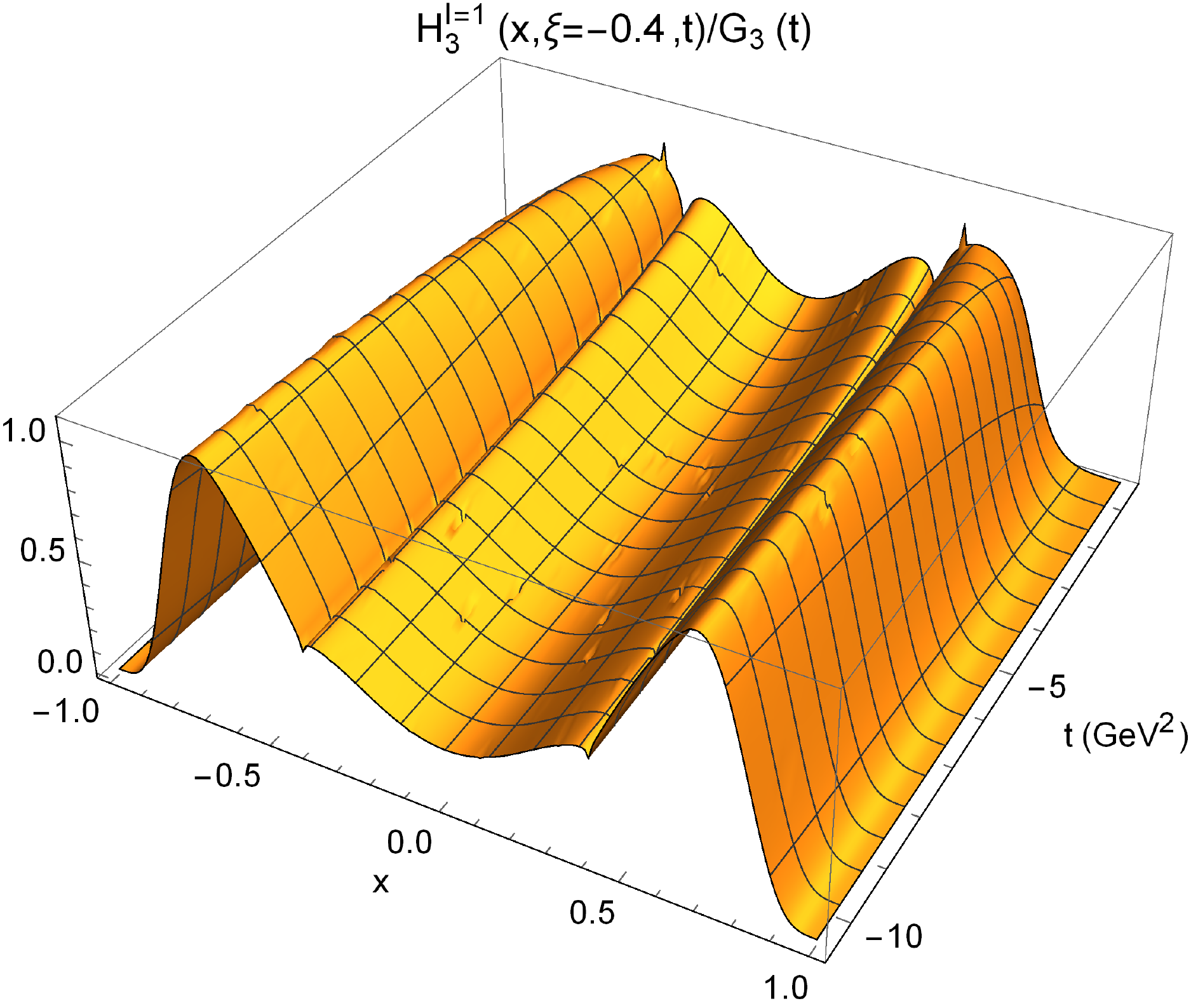}}
\caption{$\rho^+$ GPD $H_3$. The same line code is used in Fig.~\ref{fig:h1:3d}.} \label{fig:h3:3d}
\end{figure*}

\begin{figure*}[!ht]
\centering
\subfigure[$t=-0.5~\gev^2,\;\xi=0,\;-0.2,\;-0.4$]
{\label{fig:h1:t05}\includegraphics[width=7.0cm]{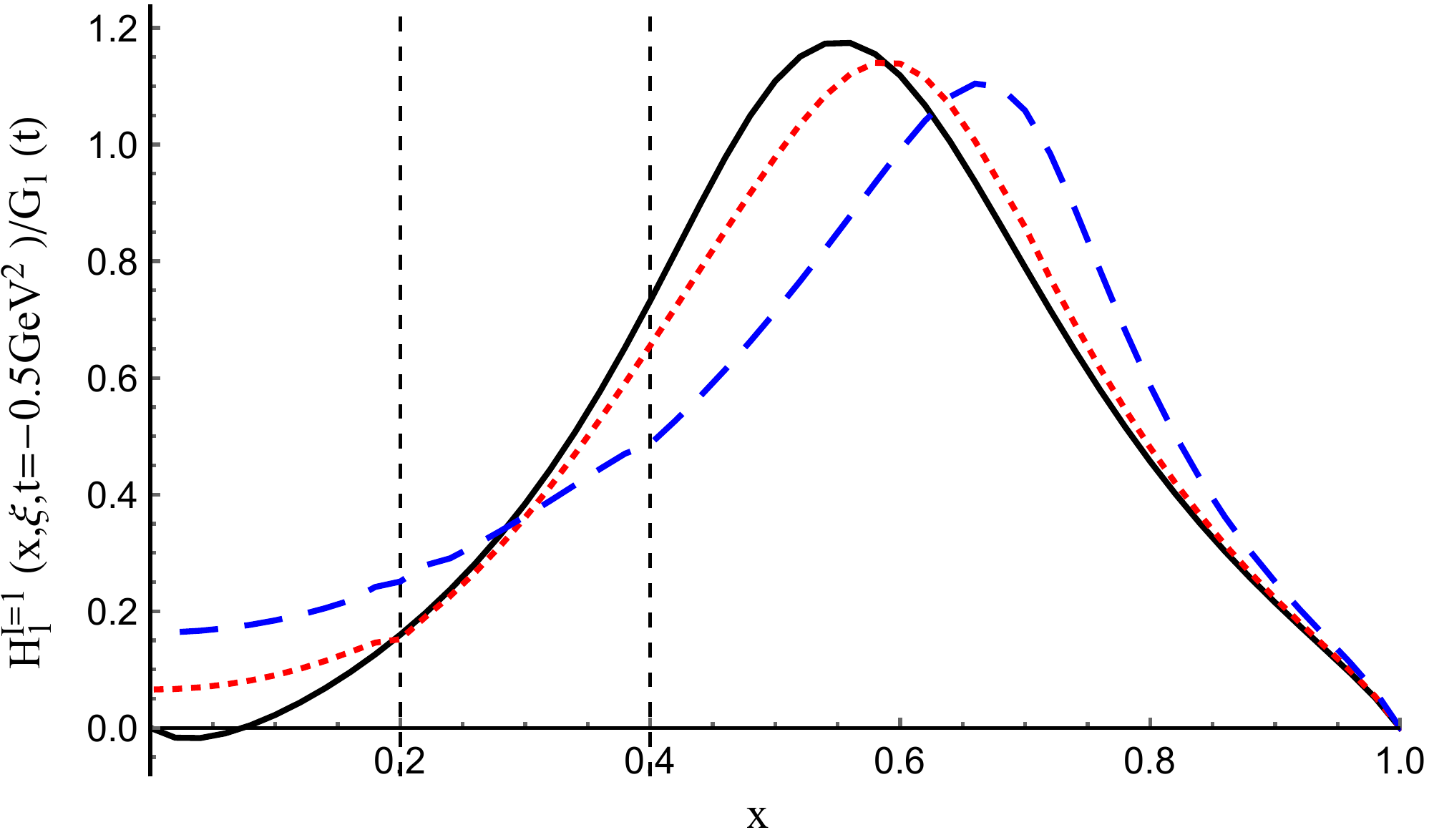}}
\subfigure[$t=-10~\gev^2,\;\xi=0,\;-0.2,\;-0.4,\;-0.6$]
{\label{fig:h1:t10}\includegraphics[width=7.0cm]{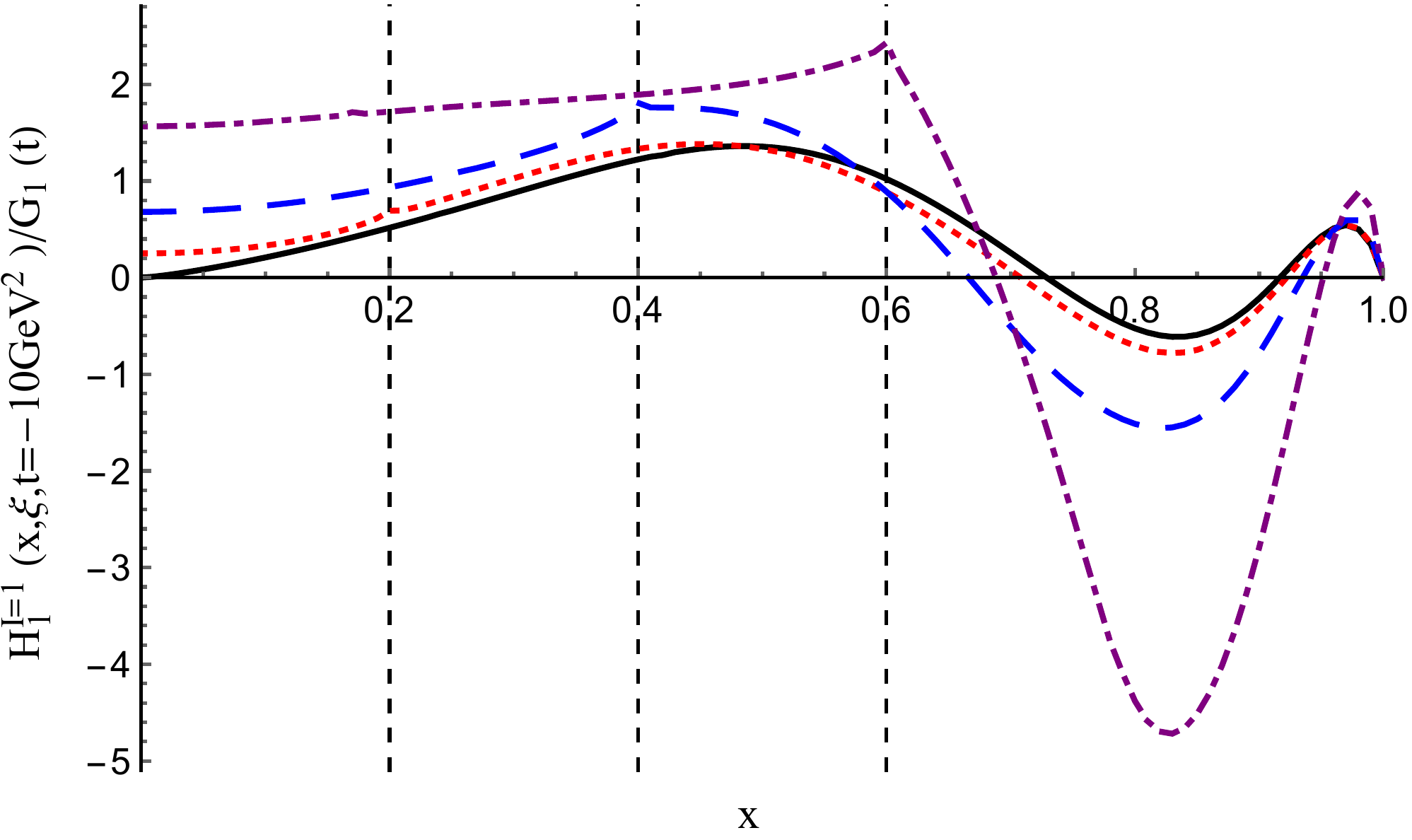}}
\caption{$\rho^+$ GPD $H_1(x,\xi, t)$ at (a) $t=-0.5~\rm{GeV}^2$ and (b) $-10~\rm{GeV}^2$. The solid black, dotted red and dashed blue curves stand for the $H_1$ with $\xi=0, -0.2$, and $-0.4$, respectively. The dotted-dashed purple curve in (b) is for $\xi=-0.6$.
    The vertical dashed lines on the x axis represent $x=\mid\xi\mid$. } \label{fig:h1}
\end{figure*}

\begin{figure*}
\centering
\subfigure[$t=-0.5~\gev^2,\;\xi=0,\;-0.2,\;-0.4$]
{\label{fig:h2:t05}\includegraphics[width=7.0cm]{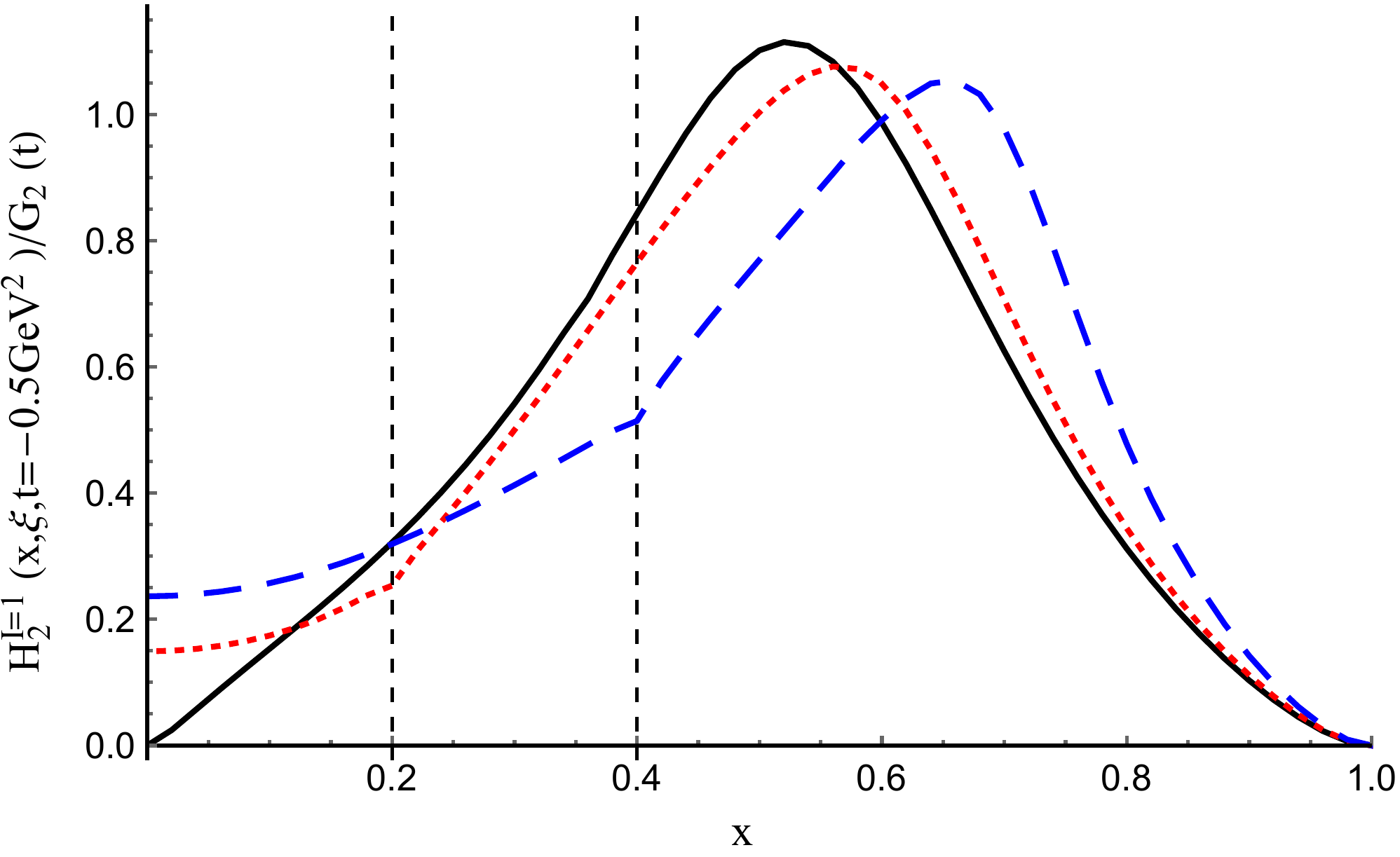}}
\subfigure[$t=-10~\gev^2,\;\xi=0,\;-0.2,\;-0.4,\;-0.6$]
{\label{fig:h2:t10}\includegraphics[width=7.0cm]{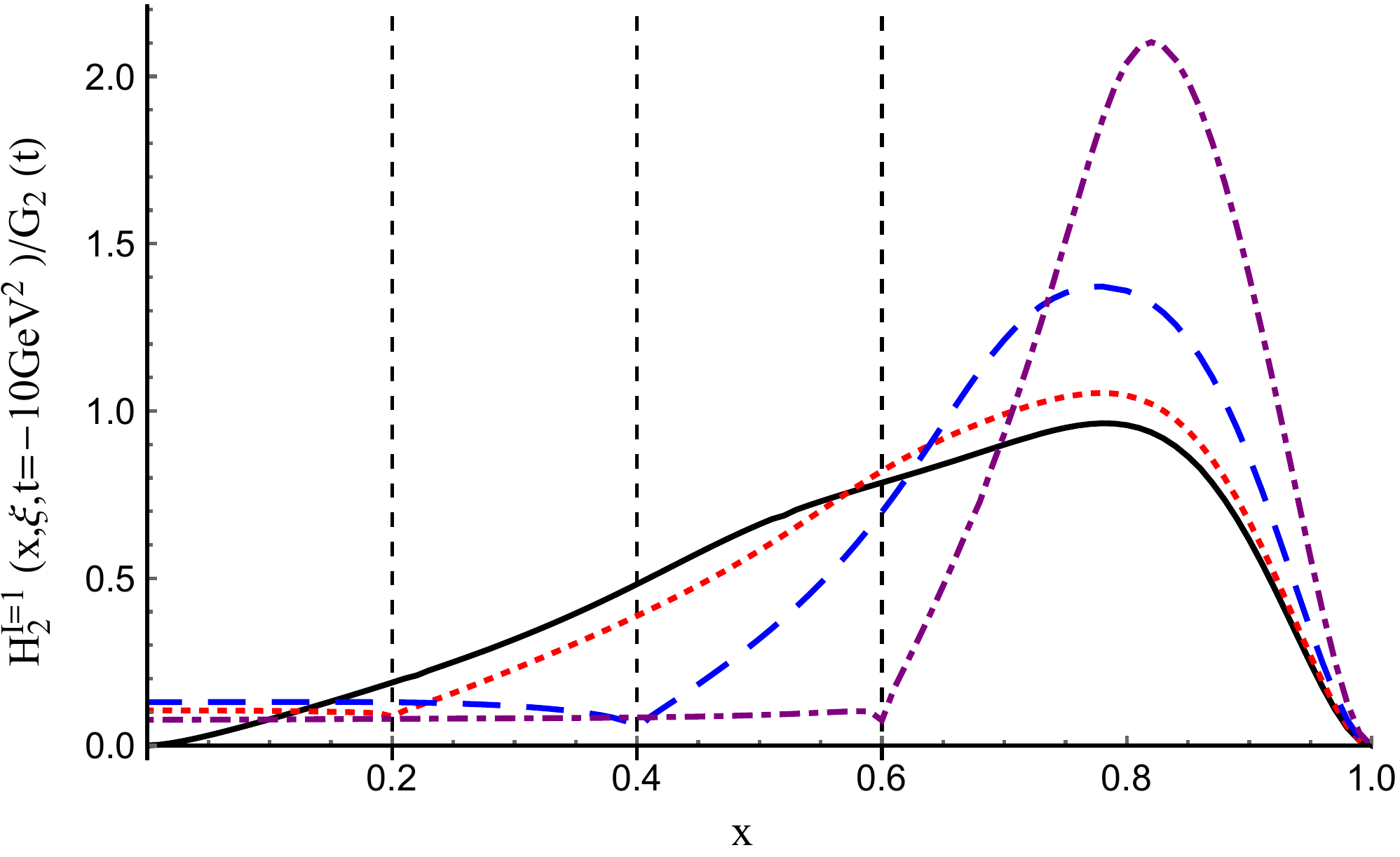}}
\caption{$\rho^+$ GPD $H_2$. The same line code is used in Fig.~\ref{fig:h1}. } \label{fig:h2}
\end{figure*}

\begin{figure*}
\centering
\subfigure[$t=-0.5~\gev^2,\;\xi=0,\;-0.2,\;-0.4$]{\label{fig:h3:t05}
\includegraphics[width=7.0cm]{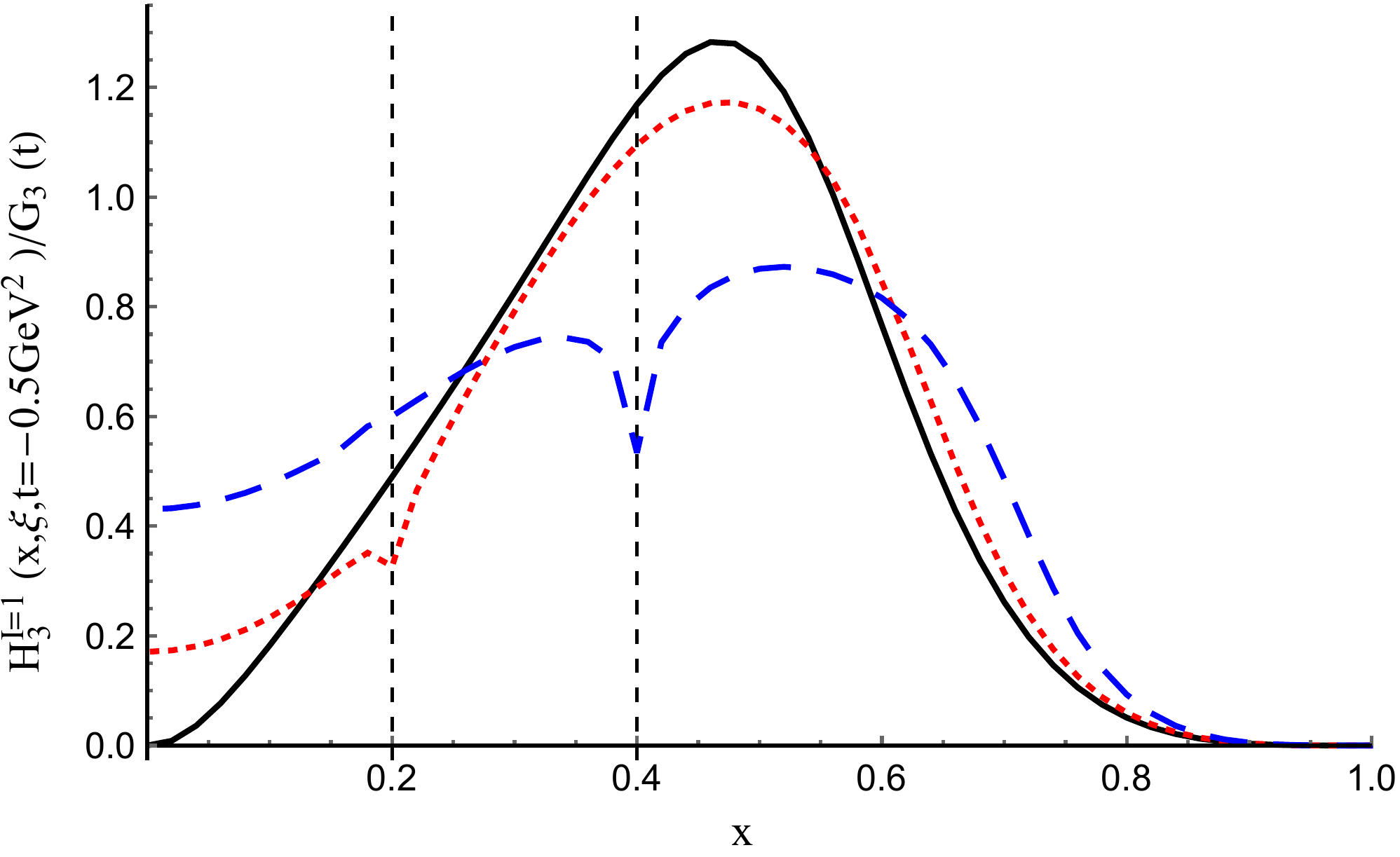}}
\subfigure[$t=-10~\gev^2,\;\xi=0,\;-0.2,\;-0.4,\;-0.6$]{\label{fig:h3:t10}
\includegraphics[width=7.0cm]{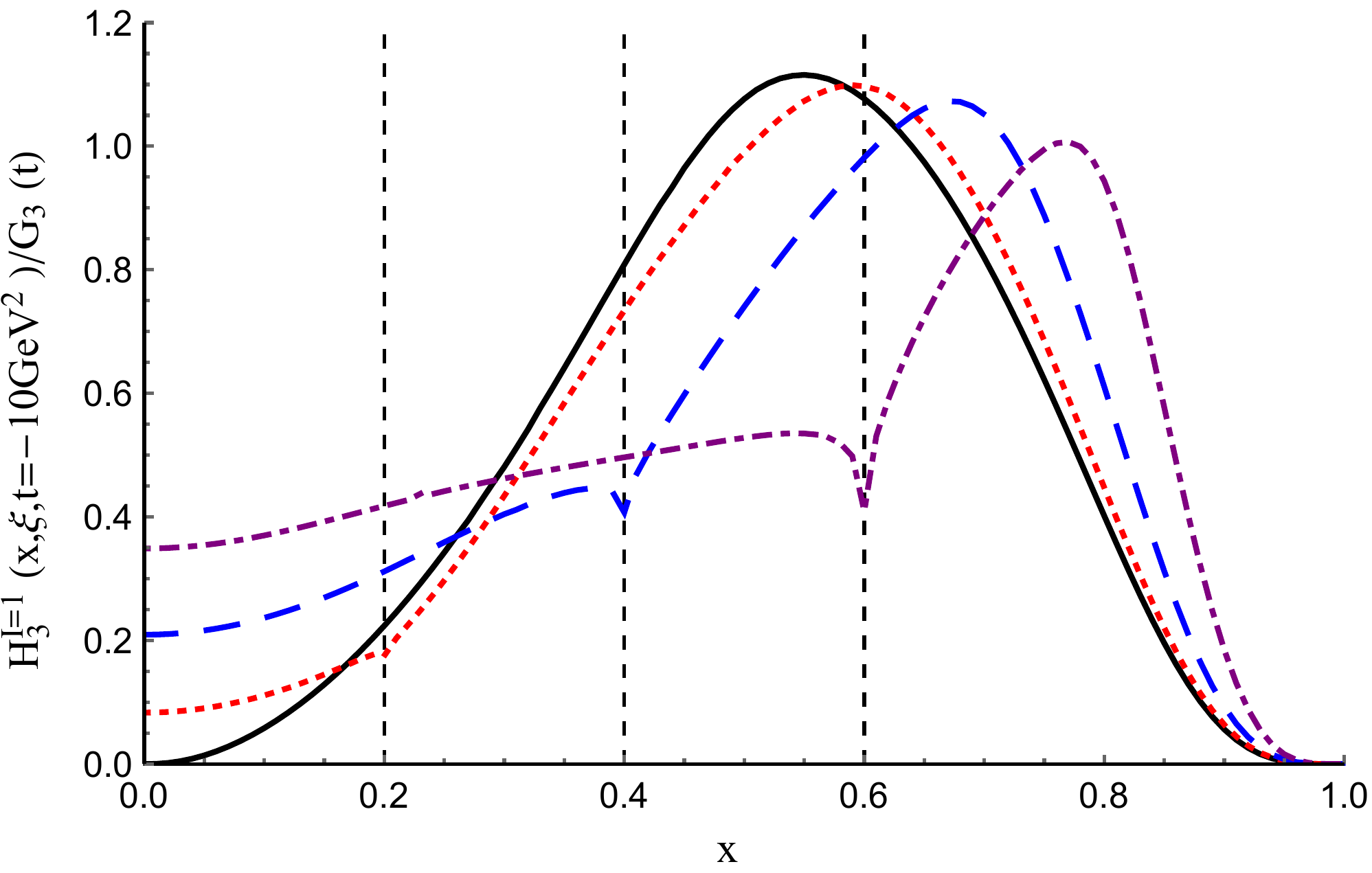}}
\caption{$\rho^+$ GPD $H_3$. The same line code is used in Fig.~\ref{fig:h1}.} \label{fig:h3}
\end{figure*}

\begin{figure}[!t]
\centerline{\includegraphics[width=8.6cm]{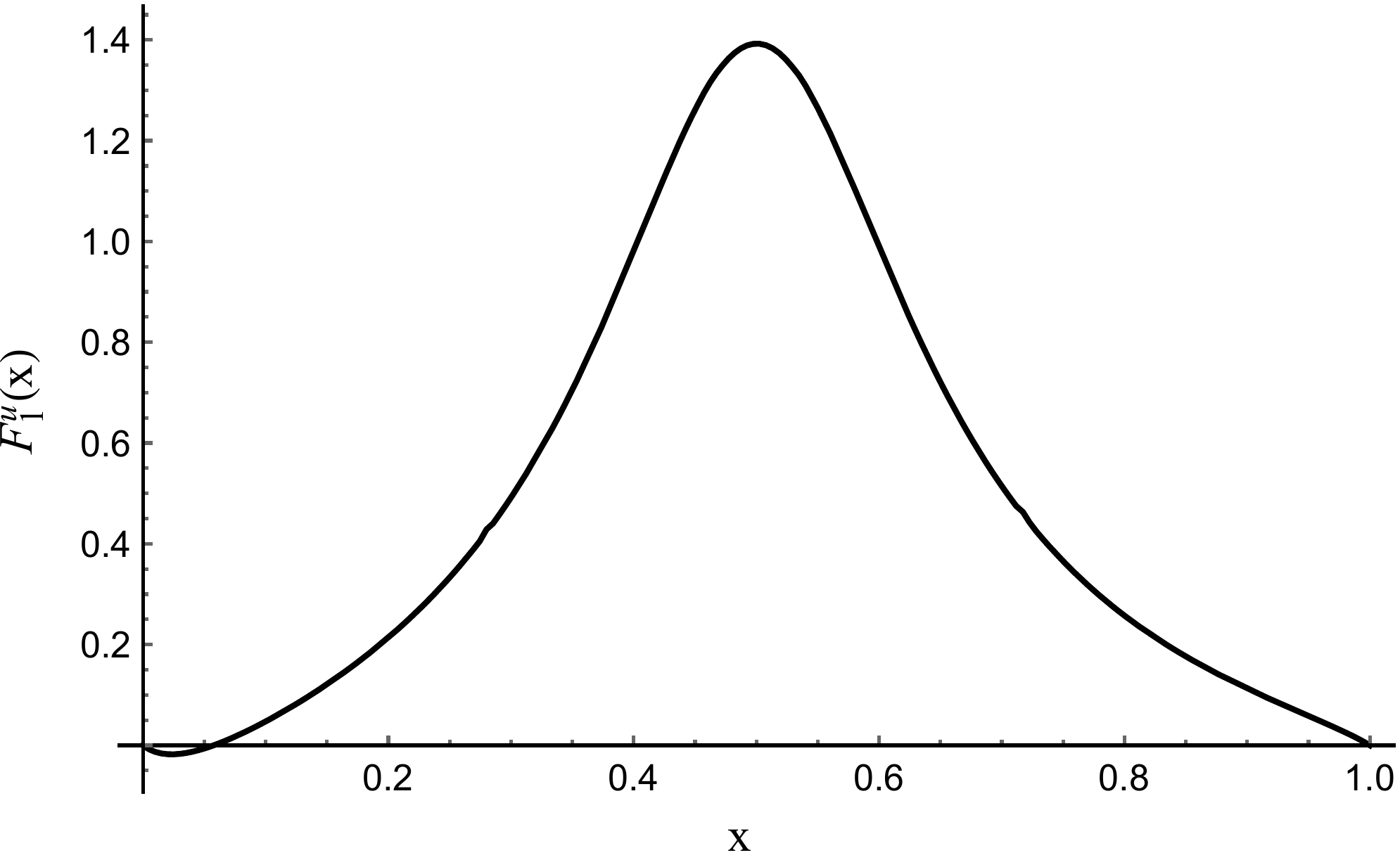}}
\caption{\label{F1} The DIS structure function $F_1$.}
\end{figure}

\begin{figure}[h]
\centerline{\includegraphics[width=8.6cm]{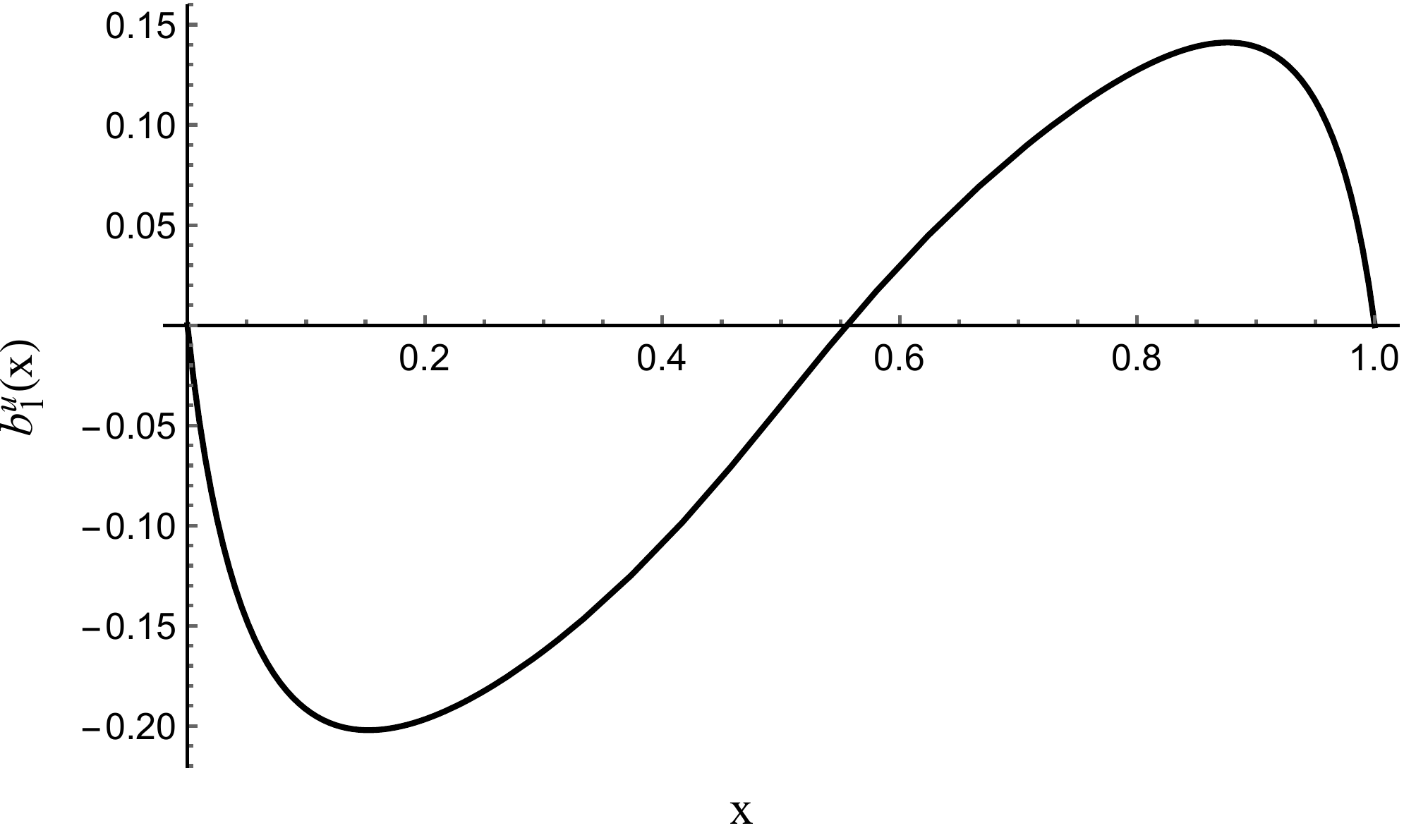}}
\caption{\label{b1} The DIS structure function $b_1$.}
\end{figure}

The calculated $\rho$ meson FFs and low-energy observables are shown in Fig.~\ref{FFs} and Table~\ref{loweo}.
The nonvalence contributions to FFs $G_{1,2,3}$ at $\xi$=$-0.2$, $-0.4$, and $-0.6$ are shown in Fig.~\ref{fig:XiTGV}. Due to the constraint $\abs{\xi}\leqslant 1/\sqrt{1-4M^2/t}$, the corresponding ${\abs t}_{min}$ are $0.10,~0.45$, and $1.33~\gev^2$, respectively. Figures~\ref{fig:h1:3d}-\ref{fig:h3:3d} show the 3D plots of the unpolarized $\rho$ meson GPDs $H_{1,2,3}$ as the functions of variables $x$ and $t$ at the two different skewnesses $\xi=0$ and $-0.4$. The values are normalized to the corresponding FFs $G_i(t)$ for a convenience of the comparison. Figures~\ref{fig:h1}-\ref{fig:h3} show $H_{1,2,3}$ at specific momentum transfers $t$ ($-0.5$ and $-10~\gev^2$) and different skewnesses $\xi$ ($0$, $-0.2$, $-0.4$, and $-0.6$). Due to the symmetry, only the $0~<~x~<~1$ regime is plotted in Figs.~\ref{fig:h1}-\ref{fig:h3}. Moreover, the two obtained structure functions, $F_1^u(x)$ and $b_1^u(x)$, are plotted in Figs.~\ref{F1} and \ref{b1}, respectively.\\

For the charge form factor $G_C$  in Fig.~\ref{FFs}, we found it has a crossing point near $t=-3.8~\gev^2$.
Moreover, the tendencies of $t$ dependence of all three obtained FFs agree with the previous results,
such as Refs.~\cite{Cardarelli1995,Choi2004,Biernat2014}. In Table~\ref{loweo}, other results of
the LCCQMs, of the point form, of the lattice QCD, and of the experiment measurement are also listed for a comparison.
Although the LCCQM proposed in the present work is inspired by former ones~\cite{Jaus2003,Choi2004} (for the meson vertex) and Ref.~\cite{Frederico2009}, etc. (for the cutoff function), different values of the model parameters, $m$ and $m_R$, are adopted here. Thus, the calculated results are different from theirs. Our calculated value of the magnetic dipole moment, $\mu=2.06$, is very close to the nonrelativistic value ($\mu=2$)~\cite{Jaus2003} and to the experimental data. In addition, the estimated mean square radius $<r^2>$ and quadrupole moment $Q_{\rho}$ in our calculation are also compatible with other calculations. It is expected that the future  measurements for the
$\rho$ meson radius and quadrupole moment may provide a test for different model calculations.\\

The Lorentz invariance requires that the FFs $G_i$ in Eq~(\ref{eq:Imm}) are frame independent, since the integration over $x$ removes the influence of different light-cone direction $n$ and therefore the integral remains $\xi$ independent. However, it is still interesting to investigate the nonvalence contribution (at $\xi\ne 0$) to  $G_i$. As one can see from Fig.~\ref{fig:XiTGV}, for all three FFs,
the valence contributions are dominant in small skewness $\abs\xi$, and the percentage of the nonvalence contributions increases as $\abs\xi$ does, which is same as the pion case~\cite{Choi2001}. As $\abs t$ increases, the nonvalence contribution in $G_1$ increases, while those in $G_2$ and $G_3$ go oppositely. Especially for $G_2$, the decrease is very distinct. It is clearly illustrated in Fig.~\ref{fig:h2:t10} for the GPD $H_2$ at $t=-10~\gev^2$. In contrast, in the pion case~\cite{Choi2001}, the nonvalence contribution to the pion form factor is especially large in the large $\abs\xi$ and small $\abs t$ region. In all three $\rho$ FFs $G_i$, we find that the sum of the numerical result of the valence and nonvalence contributions only has negligible variation over $\xi$. Thus, the frame independence of our model calculation is well satisfied. \\

\begin{table*}
\caption{\label{loweo} The $\rho$ meson low-energy observables of the mean square charge radius ($<r^2>$), the magnetic dipole ($\mu$) and the quadrupole ($Q_\rho$) moments in the units of fm${}^2$, $1/2M$, and $1/M^2$, respectively.
The results of other LCCQMs~\cite{Melo1997,Cardarelli1995,Jaus2003,Choi2004,Mello2015}, of the point-form formalism~\cite{Biernat2014}, of the lattice QCD~\cite{Owen2015}, and of the experiment measurement~\cite{Gudino2014} are also
displayed for a comparison. }
\begin{center}
	\begin{tabular}{c|cccccccccc}
	\hline
	\hline
		 &\cite{Melo1997} &\cite{Cardarelli1995} &\cite{Choi2004} &\cite{Jaus2003} &\cite{Mello2015} &\cite{Biernat2014} &\cite{Owen2015} &\cite{Gudino2014} &this work	\\\hline
	$<r^2> $	      &0.37    &0.35    &--       &--       &0.268     &--      &0.670(68)    &--          &0.52 \\\hline
	$\mu$             &2.14    &2.26    &1.92     &1.83     &2.21      &2.2     &2.613(97)    &2.1(5)      &2.06 \\\hline
	$Q_\rho$          &$-0.79$ &$-0.37$ &$-0.43$  &$-0.33$  &$-0.882$  &$-0.47$ &$-0.733(99)$ &--          &$-0.323$ \\\hline
	\hline
	\end{tabular}
\end{center}
\end{table*}



It should be stressed that our results, shown in Figs.~\ref{fig:h1:3d}-\ref{fig:h3}, are continuous at $x=\xi$ (or $\abs\xi$) as discussed earlier. The $\xi$ trajectory limits
that $-0.42~\le~\xi~\le~0$ for $t=-0.5~\gev^2$, and $-0.90~\le~\xi~\le~0$ for $t=-10~\gev^2$. As one can see from  Figs.~\ref{fig:h1:t05} and \ref{fig:h2:t05}, in small $\abs t$ region, the transition from the valence to novalence
regimes in $H_{1}$ and $H_{2}$ is quite smooth. However, in the large $\abs t$ region, as shown in Figs.~\ref{fig:h1:t10} and \ref{fig:h2:t10}, both $H_{1}$ and $H_{2}$ become sensitive to the transition in the nonzero skewness case, while Figs.~\ref{fig:h3} and \ref{fig:h3:3d} show that $H_{3}$ is very sensitive when $x\rightarrow\xi$
(or $\abs\xi$) as $\xi~\ne~0$, in both the small and large $\abs t$ regions. \\

We know that, in the forward limit ($t\rightarrow0$) and in the deep inelastic region, $H_{1}^q(x,0,0)/2$ corresponds
to the single flavor structure function $F_1^q(x)$ and $H_{5}^q(x,0,0)$ corresponds to the structure function $b_1^q(x)$. The two obtained  functions $F_1^u$ and $b_1^u$ are plotted in Figs.~\ref{F1} and \ref{b1}, respectively.
Our result for $F_1^u$ has a crossing near $x=0$, which is beyond the expectation, since as $x\rightarrow0$, $F_1^u$ should decrease to zero smoothly. This may be due to the fact that the contribution of the gluon GPDs becomes more important in the small-$x$ regime~\cite{Diehl2003}, which is beyond the scope of the present calculation. As for $b_1^u(x)$ or more general $H_{5}^u(x,\xi,t)$, the sum rules Eq.~(\ref{eq:sumrule}) requires the integral over $x$ vanishes for any $\xi$ and $t$. Our numerical result holds the sum rule for $H_4^u$ quiet well, but for $H_5^u$, the integral deviates from zero by, at most, $\sim~6.5~\%$ [with repect to $G_C(0)=1$] when $-7~\gev^2\le t \le 0$. The violation of sum rules of $H_4$ and $H_5$ is also encountered in the deuteron case, such as the numerical model in Ref.~\cite{Cano2004}. In addition, the symmetry around $x\sim1/2$ preserves approximately for both $F_1^u(x)$ and $b_1^u(x)$ in our phenomenological model calculation. This symmetry conforms to the isospin and crossing symmetries, which reduces $u_{\rho^+}{(x)}={\bar{d}}_{\rho^+}{(1-x)}$.\\

\section{QCD evolution}\label{sec:evolution}

It is known that the low-energy chiral quark model provides the initial conditions for the QCD evolution. The present work assumes that the valence quarks carry all the momentum at a factorization scale $Q_0$. To compare our result with the lattice calculation, the evolution is needed. As far as we know, Ref.~\cite{Best1997} is the only one lattice QCD calculation for the moments of the unpolarized $\rho$ meson, which is at the scale $Q= 2.4~\gev$ with quenched approximation. Its results are supported by the latter QCD sum rule calculation in Ref.~\cite{Oganesian2001}. The leading order(LO) DGLAP evolution for the moments of the single flavor structure function $F_1^u(x)$ reads
\eq
\label{eq:dglap}
\frac{V_n^u(Q)}{V_n^u(Q_0)} = \left( \frac{\alpha(Q)}{\alpha(Q_0)} \right)^{\gamma_n^{(0)}/(2\beta_0)}
\en
where the valence-quark momentum fractions $V_n^u=M_{n+1}\left[H_{1}^u(x,0,0)\right]=2M_{n+1}\left[F_1^u(x)\right]\sim a_{n+1}$ and
the running coupling constant is
\eq
\alpha(Q) &=& \frac{4\pi}{\beta_0 \text{log}(Q^2/\Lambda^2_{QCD})} \ , 
\en
where $\beta_0 = {11N_c}/{3}  - {2N_f}/{3} $ with $N_c=N_f=3$ and
\eq
\Lambda_{QCD} = 0.226~GeV 
\en
being employed~\cite{Broniowski2008,Broniowski200878}. The percentage of the $\rho$ total momentum carried by
the valence quarks is $V_1=V_1^u +V_1^{\bar{d}}$. In Ref.~\cite{Best1997}, it was found that $V_1^q(Q=2.4~\gev)=0.33(2)$ and therefore $V_1$ is about $70\%$. At the quark model point $Q_0$, i.e. the model factorization scale, $V_1(Q_0)$ turns out to be
\eq
V_1(Q_0)=1 , \quad G_1(Q_0)+S_1(Q_0)=0 ,
\en
from the downward LO DGLAP evolution. In the above equation, $G_1(Q)$ and $S_1(Q)$ are the gluon and sea momentum fractions, respectively. Thus, quark model point $Q_0$ is
\eq \label{eq:scaleQ0}
Q_0 = 528_{-62}^{+77} \; \text{MeV} \ .
\en
The error bars come from the uncertainty in lattice result for $V_1^q$.
This is a rather low scale, however, the typical expansion parameter $\alpha(Q_0)/(2\pi)=0.131_{-0.023}^{+0.018}$
makes the perturbation theory meaningful. It should be mentioned that our numerical result for $V_1$
is $1.02$, which diverges from unity by 2\%.  At the scale $Q=2.4~\gev$, the results for $V_{1,2,3}$ (or $a_{2,3,4}$) and $d_{2,3,4}$ of Ref.~\cite{Best1997} are
\eq
V_1^u=0.33(2)\ , \quad V_2^u=0.17(5)\ , \quad V_3^u=0.06(4)\ , \\
d_2=0.29_{-23}^{+22}\ , \quad d_3=-0.001(15)\ , \quad d_4=-0.01(6) \ .
\en
After the LO DGLAP evolution to the lattice scale, our model predicts
\eq
V_1^u=0.34(2)\ , \quad V_2^u=0.15(1)\ , \quad V_3^u=0.08(1)\ , \\
d_2=0.044(3)\ , \quad d_3=0.048(5)\ , \quad  d_4=0.039(5)\ ,
\en
where the error bars came from the uncertainty of the predicted scale $Q_0$ in Eq.~(\ref{eq:scaleQ0}).
As one can see, except for $d_2$, the lowest moments of $b_1^q(x)$, our results agree well with the lattice calculations.
However, it should be emphasized that Ref.~\cite{Best1997} concluded their
$d_2$ value being ``surprisingly" large at the scale of $Q=2.4~\gev$, since $b_1^q(x)$ should vanish if the $\rho$ meson is in a pure ${}^3S_1$ state. We believe that our estimated smaller value
for $d_2$ is more reasonable since only the ${}^3S_1$-wave coupling is taken into account in the present
calculation and the ${}^3D_1$ admixture is thought to be small ($\sim 1\%$) as mentioned earlier.

\section{Summary and conclusions}
\label{sec:summary}

In the present work, we perform a calculation for the $\rho$ meson unpolarized GPDs by employing a light-cone constituent quark model and using the isospin combination~\cite{Broniowski2008,Broniowski200878}. The smeared $\rho-q\bar{q}$ meson vertex, which represents the nonperturbative QCD effect, is adopted following Ref.~\cite{Choi2004} but using the symmetric loop momentum convention in order to satisfy the isospin symmetry. The three $\rho$ meson FFs and some other low-energy observables are calculated. Our results are compatible with the previous calculations. By considering the sum rules of GPDs, the unpolarized DIS structure functions have also been estimated, and the symmetric distribution is basically maintained
in our numerical calculation. This feature reflects the isospin and crossing symmetries. At $x=\xi$ (or $\abs\xi$) where the DGLAP and ERBL regimes meet, GPDs in our calculation are continuous, as required by the consistency
of factorization at leading twist~\cite{Diehl2003}. After the QCD evolution, the model predictions for the moments
of structure functions are compared with the lattice calculation.
The obtained factorization scale $Q_0$ is a rather low one
in this work. However, the corresponding typical expansion parameter is still small enough to make the
perturbative calculation meaningful.
It is encouraging that all the first three-order moments in our calculation
are compatible with the lattice calculation at the same scale ratio. The present model can be also applied for
the polarized GPDs of the $\rho$ meson, and such a calculation is in a progress. Moreover, a calculation for the deuteron GPDs is also expected in the future.\\

\section*{Acknowledgements}

We would like to thank Yu Jia, Wei Sun, Yu Lu, and B. Pire for their useful and constructive discussions.
This work is supported by the National Natural Sciences Foundations of China under Grant
No. 11475192 and No. 11521505, and by the fund provided to the Sino-German CRC 110 "Symmetries
and the Emergence of Structure in QCD" project by NSFC under Grant No. 11621131001.

\section*{APPENDIX: Extracting the unpolarized GPDs.}
\label{sec:appendix}

The following is the method to extract the unpolarized GPDs $H_i$. First, the loop
integral, Eq.~(\ref{Vlf}), after excluding the two polarization vectors $\ep_{\mu}$ and $\ep'^*_{\nu}$, is
\eq
V^{u\, ; \mu\nu}_{ } (x, \xi, t)
  &=&  \frac{M^2}{f_\rho^2}
  \frac{ 1 }{ 2(2\pi)^3 \sqrt{\omega_{p'}\omega_{p} } }
  \int \frac{d^4 k  }{ (2\pi)^4 }
  \, \delta \left[ x P^+ -k^+ \right]
  \nonumber \\  &&
  \times (-) Tr \Bigg\{
  \frac{\imath ( \slash{k}-\slash{P}+m ) }{ (k-P)^2-m^2 + \imath \epsilon}
  \gamma^\nu
  \frac{\imath ( \slash{k}+\frac{\slash{\Delta}}{2} +m ) }{ (k+\frac{\Delta}{2})^2 -m^2 + \imath \epsilon}
  \slash{n}
  \frac{\imath ( \slash{k}-\frac{\slash{\Delta}}{2} +m ) }{ (k-\frac{\Delta}{2})^2 -m^2 + \imath \epsilon}
  \gamma^\mu
  \Bigg\}
  \nonumber \\  &&
  \times \frac{c }{ [ (k-P)^2-m^2_R+ \imath \epsilon] [ ({k}+\frac{\Delta}{2})^2-m^2_R+ \imath \epsilon] }
  \times \frac{c }{ [ (k-P)^2-m^2_R+ \imath \epsilon] [ ({k}-\frac{\Delta}{2})^2-m^2_R+ \imath \epsilon] }
  \nonumber \\ [1.3em]
  &=& -g^{\mu \nu} H^u_{1} + \frac{n^\mu P^\nu + P^\mu n^\nu }{n \cdot P } H^u_{2} - \frac{ 2 P^\mu P^\nu}{ M^2}H^u_{3}
    + \frac{n^\mu P^\nu - P^\mu n^\nu }{n \cdot P } H^u_{4} + \bigg\{ \frac{ M^2 n^\mu n^\nu}{(n \cdot P )^2} + \frac{1}{3} g^{\mu \nu} \bigg\} H^u_{5}   \ ,
\label{eq:VectorGPDs}
\en
Then, by contracting with the five tensors, one gets  five independent equations as
\eq
\left(
\begin{array}{c}
g_{\mu v} \\ n_\mu n_\nu \\ n_\mu P_\nu \\ n_\nu P_\mu \\ P_\mu P_\nu
\end{array}
\right)
\cdot V^{u\, ; \mu\nu}=
\left(
\begin{array}{ccccc}
 -4 & 2 & -\frac{2 P^2}{M^2} & 0 & \frac{4}{3} \\
 0 & 0 & -\frac{2 ( n \cdot  P )^2}{M^2} & 0 & 0 \\
 -(n \cdot  P) & n \cdot  P & -\frac{2 (n \cdot  P) P^2}{M^2} & -(n \cdot  P) & \frac{n \cdot  P}{3} \\
 -(n \cdot  P) & n \cdot  P & -\frac{2 (n \cdot  P) P^2}{M^2} & n \cdot  P & \frac{n \cdot  P}{3} \\
 -P^2 & 2 P^2 & -\frac{2 P^4}{M^2} & 0 & M^2+\frac{P^2}{3} \\
\end{array}
\right)
\cdot
\left(
\begin{array}{c}
H^u_1 \\ H^u_2 \\ H^u_3 \\ H^u_4 \\ H^u_5 \\
\end{array}
\right).
\en
Finally, the explicit expressions for GPDs are obtained:
\eq
\left(
\begin{array}{c}
H^u_1 \\ H^u_2 \\ H^u_3 \\ H^u_4 \\ H^u_5 \\
\end{array}
\right)
=
\left(
\begin{array}{ccccc}
 \frac{1}{6} \left(\frac{{P}^2}{M^2}-3\right) & \frac{{P}^2 \left({P}^2-M^2\right)}{2 M^2 \left( {n} \cdot  {P} \right)^2} & \frac{M^2-{P}^2}{2 M^2 \left({n} \cdot  {P}\right)} & \frac{M^2-{P}^2}{2 M^2 \left({n} \cdot  {P}\right)} & \frac{1}{3 M^2} \\
 -\frac{1}{2} & -\frac{3 {P}^2}{2 \left( {n} \cdot  {P} \right)^2} & \frac{1}{{n} \cdot  {P}} & \frac{1}{{n} \cdot  {P}} & 0 \\
 0 & -\frac{M^2}{2 \left( {n} \cdot  {P} \right)^2} & 0 & 0 & 0 \\
 0 & 0 & -\frac{1}{2 \left({n} \cdot  {P}\right)} & \frac{1}{2 \left({n} \cdot  {P}\right)} & 0 \\
 \frac{{P}^2}{2 M^2} & \frac{3 {P}^4}{2 M^2 \left( {n} \cdot  {P} \right)^2} & -\frac{3 {P}^2}{2 M^2 \left({n} \cdot  {P}\right)} & -\frac{3 {P}^2}{2 M^2 \left({n} \cdot  {P}\right)} & \frac{1}{M^2} \\
\end{array}
\right)
\cdot
\left(
\begin{array}{c}
g_{\mu v} \\ n_\mu n_\nu \\ n_\mu P_\nu \\ n_\nu P_\mu \\ P_\mu P_\nu
\end{array}
\right)
\cdot V^{u\, ; \mu\nu}
\en
\bibliography{ref0716}
\end{document}